\documentclass[prb,reprint,notitlepage,superscriptaddress,twocolumn,floatfix,aps]{revtex4-2}
\usepackage{amsmath}
\usepackage{graphicx}
\usepackage[caption=false]{subfig}
\usepackage{lmodern}
\usepackage{amsmath}
\usepackage{physics}
\usepackage{natbib}
\usepackage{tikz}
\usepackage{color}
\usepackage{diagbox}
\usepackage{bm}
\usepackage{amssymb}
\usepackage{tabularx}
\usepackage{hyperref}
\usepackage{tabularx}
\usepackage{bbold}
\usepackage[normalem]{ulem}
\hypersetup{
    pdftoolbar=true,        
    pdfmenubar=true,        
    pdffitwindow=false,     
    pdfstartview={FitH},    
    pdftitle={My title},    
    pdfauthor={Author},     
    pdfsubject={Subject},   
    pdfcreator={Creator},   
    pdfproducer={Producer}, 
    pdfkeywords={keyword1} {key2} {key3}, 
    pdfnewwindow=true,      
    colorlinks=true,       
    linkcolor=brickred,          
    citecolor=bleu,        
    filecolor=magenta,      
    urlcolor=pblue           
}
\definecolor{pblue}{rgb}{0.0, 0.4, 0.65}
\definecolor{frenchblue}{rgb}{0.0, 0.45, 0.73}
\definecolor{brickred}{rgb}{0.8, 0.25, 0.33}
\definecolor{bleu}{rgb}{0.0, 0.5, 0.69}

\newcommand{\beq} {\begin{equation}}
\newcommand{\eeq} {\end{equation}}
\newcommand{\bea} {\begin{eqnarray}}
\newcommand{\eea} {\end{eqnarray}}
\newcommand{\be} {\begin{equation}}
\newcommand{\ee} {\end{equation}}
\renewcommand{\(}{\left(}
\renewcommand{\)}{\right)}
\renewcommand{\[}{\left[}
\renewcommand{\]}{\right]}

\newcommand{\oncite}[1]{Ref.\ [\onlinecite{#1}]}
\definecolor{ncs}{rgb}{0.0, 0.33, 0.74}

\DeclareMathOperator{\sgn}{sgn}

\DeclareMathOperator{\Imm}{Im}

\newcommand{\ms}{\mathsf}
\newcommand{\bs}{\boldsymbol}
\newcommand{\mc}{\mathcal}

\newcommand{\darkred}[1]{{\color[rgb]{0.6,0,0}{#1}}}

\definecolor{blu}{rgb}{0.87, 0.36, 0.51}

\begin{document}

\title{Higher-order topological superconductors from Weyl semimetals}

\author{Ammar Jahin}
\affiliation{
 Department of Physics, University of Florida, 2001 Museum Rd, Gainesville, FL 32611
 }
\author{Apoorv Tiwari}
\affiliation{Department of Physics, University of Zurich, Winterthurerstrasse 190, 8057 Zurich, Switzerland}\affiliation{Condensed Matter Theory Group, Paul Scherrer Institute, CH-5232 Villigen PSI, Switzerland}
\author{Yuxuan Wang}
\affiliation{
 Department of Physics, University of Florida, 2001 Museum Rd, Gainesville, FL 32611
}

\date{\today}

\begin{abstract}
We propose that doped Weyl semimetals with four Weyl points are natural candidates to realize higher-order topological superconductors, which exhibit a fully gapped bulk while the surface hosts robust gapless chiral hinge states. We show that in such a doped Weyl semimetal, a featureless finite-range attractive interaction favors a $p+ip$ pairing symmetry. By analyzing its topological properties, we identify such a chiral pairing state as a higher-order topological superconductor, which depending on the existence of a four-fold roto-inversion symmetry $\ms{R}_{4z}$, is either intrinsic (meaning that the corresponding hinge states can only be removed by closing the bulk gap, rather than modifying the surface states) or extrinsic. We achieve this understanding via various methods recently developed for higher-order topology, including Wannier representability, Wannier spectrum, and defect classification approaches.
 For the $\ms{R}_{4z}$ symmetric case, we provide a complete classification of the higher-order topological superconductors. We show that such second-order topological superconductors exhibit chiral hinge modes that are robust in the absence of interaction effects but can be eliminated at the cost of introducing surface topological order.

\end{abstract}

\maketitle

\tableofcontents

\section{Introduction}
{Topological superconductivity~\cite{qi-zhang-2011, Alicea_2012, sato-ando-2017, Sato_2017} combines two fascinating topics in condensed matter physics, topological phases of matter and unconventional superconductivity, and is the key component of fault-tolerant topological quantum computation~\cite{ivanov-2001, nayak-2008}. Over the past decade, significant progress has been made in classifying topological superconductors with internal and/or crystalline symmetries. For the purpose of classification, these phases are often treated as free fermion states. For experimental realizations, much of the focus has been placed on ideas similar to the Fu-Kane superconductor~\cite{fu-kane-2008} where a conventional superconductor is in proximity with a topological material. On the other hand, unconventional superconductors with nontrivial (i.e., non-$s$-wave) pairing symmetries can exhibit even richer symmetry-breaking and topological properties. 
The understanding and prediction of these unconventional topological superconductors necessarily require a synergy of band structure and electronic interaction effects. }

{The notion of band topology has recently been extended to \emph{higher-order} topology~\cite{Benalcazar-2017,benalcazar-bernevig-hughes-prb-2017,Schindlereaat-2018,Yuxuan-2018, Piet-2017, Chen-Fang-2019, Peterson_2018, Serra_Garcia_2018, Imhof_2018, Ezawa-2018, Fulga-2018, Wieder_2020, Mittal_2019, He_2020, Zeng_2020, Zhang_2020, Chen_2020, El_Hassan_2019, Fan_2019, Chen_2019, Xie_2019, zhang2019higherorder, Apoorv_2020}, with protected gapless states localized at the corners and hinges of the sample. This opens up a new avenue for novel topological superconductivity~\cite{Yuxuan-2018, Geier_2020, Skurativska_2020, Das-Sarma-2019a, Das-Sarma-2019b, Das-Sarma-2020, Roberts_2020}, where many interesting open questions abound, including  classification of such phases and its potential application in topological quantum computation. Just like regular unconventional topological superconductors, the realization of higher-order topological superconductivity via an intrinsic pairing instability typically has stringent requirements on both the normal state band structure and the pairing symmetry in an intrinsic superconductor. There have been several recent proposals along these lines, including potential higher-order topological superconducting phases (HOTSC) in FeSeTe, in two-dimensional Dirac semimetals~\cite{Yuxuan-2018,Zhongbo-2019,Zhongbo-2019b, Das-Sarma-2019a, Das-Sarma-2019b, Das-Sarma-2019c}, and in superconductors with unconventional $p+id$ pairing symmetry~\cite{Yuxuan-2018,bitan-2020}. Alternatively, it has been pointed out in several recent works~\cite{Wang-Zhong-2018,Zhang-Fan-2018} that superconducting proximity effects between a quantum spin Hall insulator and a $d$-wave superconductor also realizes a HOTSC phase.}


{In this work we show that thanks to its normal state band structure, interacting topological semimetals are natural candidates for hosting HOTSCs. A number of previous~\cite{wang-nandkishore-2017, Shapourian-2018, tiwari2020chiral,li-haldane-2018,sun-li-2020} works have shown that topological semimetals provide a promising avenue for realizing novel topological superconducting phases, including fully gapped ones and those with topologically protected nodal points.} Here we analyze the fully gapped superconudcting phase that emerges from an interacting time-reversal symmetric Weyl semimetal. A minimal model of such a system consists of two bands with four co-planar Weyl points. With a proper chemical potential within the width of Weyl bands, there exist four Fermi pockets around each Weyl point. We find that in the presence of a finite-range attractive interaction {(as opposed to an on-site or short-ranged one)}, the leading instability is toward a chiral $p$-wave order, which spontaneously breaks time-reversal symmetry. While the resulting superconductor is fully gapped in the bulk, it hosts gapless chiral Majorana modes at its hinges that are perpendicular to the plane of Weyl points. These gapless hinge states are a characteristic of second-order topology. We examine the topological properties in the presence of a four-fold rotoinversion symmetry $\ms R_{4z}$ via several different methods, including the analysis of Wannier obstruction and the defect classification approach and find that the bulk has no well-defined Wannier representation that respects all the symmetries of the system.

Using the defect classification approach that we developed for higher-order topology in an earlier work \cite{tiwari2020chiral}, we find that the defect Hamiltonian $H(\bm k, \theta)$ for a tube enclosing the hinge has a second Chern number protected by $\ms R_{4z}$ symmetry. This further confirms the robustness of the chiral hinge modes and second-order topology. Next, we extend our focus to the general class of $\ms R_{4z}$-symmetric superconductors in 3d, and obtain a full classification. We demonstrate that while the chiral hinge modes are robust for a free fermion system, they can be eliminated in the presence of strong interactions on the surface by inducing an anomalous surface topological order~\cite{Apoorv_2020}. 

We also analyze the situation in the absence of $\ms R_{4z}$ symmetry. Of important relevance to this case is a four-band time-reversal invariant Weyl semimetal. In this situation two pairs of Weyl points come from different bands that are Kramers partners, and four-fold symmetries are absent. Despite the reduced symmetry, the chiral $p$-wave pairing order remains the leading pairing channel. However, in the absence of $\ms R_{4z}$, the aforementioned classification of HOTSC does not apply. Nevertheless, we show that the chiral hinge modes remain a robust feature of the spectrum of a finite sized sample. We show this by directly solving the defect Hamiltonian corresponding to the portion of the surface around a hinge. These hinge states can be understood as coming from \emph{extrinsic} second-order topology, as they can be eliminated by modifying the surface without closing the gap in the bulk. The Wannier obstruction of the surface states remain present, consistent with the fact that the hinge modes are protected by the surface gap.

{The rest of this paper is organized as follows. In Sec.~\ref{sec:2} we introduce the model for the normal state and analysis its pairing instabilities in the presence of an attractive interaction. In Sec.~\ref{sec:roto_inv_model} we show that such a chiral $p$-wave superconductor has nontrivial second-order topology in the presence of $\ms R_{4z}$ symmetry. In Sec.~\ref{sec:class} we obtain a full classification of the higher-order topology for 3d $\ms R_{4z}$ symmetric superconudctors, and in Sec.~\ref{Sec:STO} we discuss the fate of the gapless hinge modes in the presence of strong surface interactions. In Sec.~\ref{sec:botsc} we show that the chiral hinge modes remain robust in the absence of $\ms R_{4z}$ symmetry. }

\section{Time-reversal invariant Weyl semimetal and its pairing instabilities}
\label{sec:2}

\subsection{Normal state}
Consider the following two-band lattice model for a Weyl semimetal, $H = \int d\bm k \psi^\dagger_{\bm k} \mathcal{H}_{n}(\bm k) \psi_{\bm k} $, with the single-particle Hamiltonian given by
\begin{align}
    {H}_n(\bm k) =&\;  \bm f(\bm k) \cdot \bm \sigma - \mu,
    \label{eq:general_two_band_WSM}
\end{align}
where $\sigma_i$'s are Pauli matrices acting on an internal band space.
{The Weyl nodes of the band structure are given by the condition $\bm{f}(\bm k_0) =0$, which are in general isolated points in three dimensions.} We impose a time-reversal symmetry $\ms T$ such that
\begin{gather}
    \ms T  H_n(\bm k) \ms T^{-1} =  H_n( - \bm k).
\end{gather} In general the two bands are non-degenerate other than at the Weyl points, which are not at high-symmetry points, and we take $\ms T^2 = 1$ so that the time-reversal symmetry does not enforce any Kramer's degeneracy. With no loss of generality we choose the time reversal symmetry to be,
\begin{align}\label{eq:TRS_form}
    \ms T = \mathcal{K}, 
\end{align} 
where $\mc K$ is the complex conjugation operator. Other choices are related by unitary transformations in the band basis. Time-reversal symmetry requires
\begin{gather}
    f_{1,3}(-\bm k) = f_{1,3}(\bm k), \nonumber  \\ 
    f_2(-\bm k) = -f_2(\bm k).\label{eq:TRS-constrains}
\end{gather} 
In the presence of time-reversal symmetry, there are a minimum of four Weyl points that are pairwise related. We primarily focus on this minimal case in this work. 
The pair of Weyl points related by time-reversal each carry a monopole charge (Chern number) $C=1$, while the other pair each carry $C=-1$ in accordance with the Nielson-Ninomiya theorem \cite{Nielson_ninomiya}.

\begin{figure}[t] 
    \includegraphics[scale=0.8, trim = 0 10 0 10, clip]{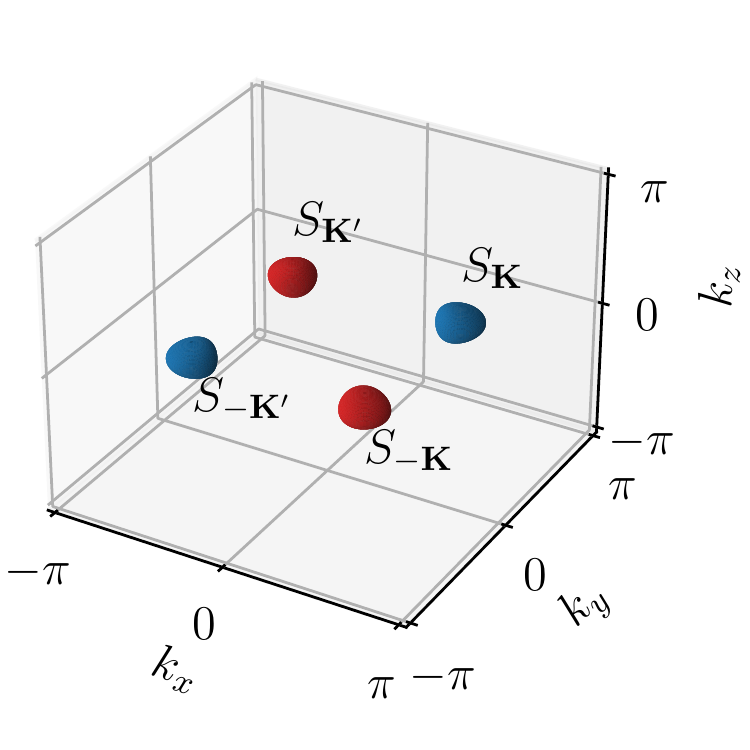}
    \caption{
     The position of the four ellipsoidal Fermi surfaces in the Brilliouin zone. The color of a Fermi surface denotes the chirality of the Weyl point it encloses with {\bf{\darkred{red}}} (resp. {\bf{\color{ncs}{blue}}}) with $C=1$ (resp. $-1$).
    }
    \label{fermisurfaces}
\end{figure}

{Additionally, we impose spatial symmetries relating all four Weyl points. While the simplest possibility would be a four-fold rotation in the plane of Weyl points, such a symmetry is incompatible with the fact that the four Weyl points carry alternating $\pm 1$ monopole charges under a four-fold rotation. Instead such a configuration of Weyl-points can be stabilized by a four-fold roto-inversion symmetry, given by a composite transformation  
$\ms R_{4z} = \ms C_{4z}\ms{M}_z$, where $\ms C_{4z}$ is a fourfold rotation around the $z$-axis and $\ms M_z$ is reflection along the $z$-axis, under which}
 \begin{gather}
    \ms R_{4z} \mc H_n(\bm k) \ms R_{4z}^{-1} = \mc H_n(\ms R_{4z} \bm k), 
\end{gather}
with $\ms R_{4z}:(k_x,k_y,k_z) \to (-k_y, k_x, -k_z)$.

\begin{figure}
    \centering
    \includegraphics[scale=0.9]{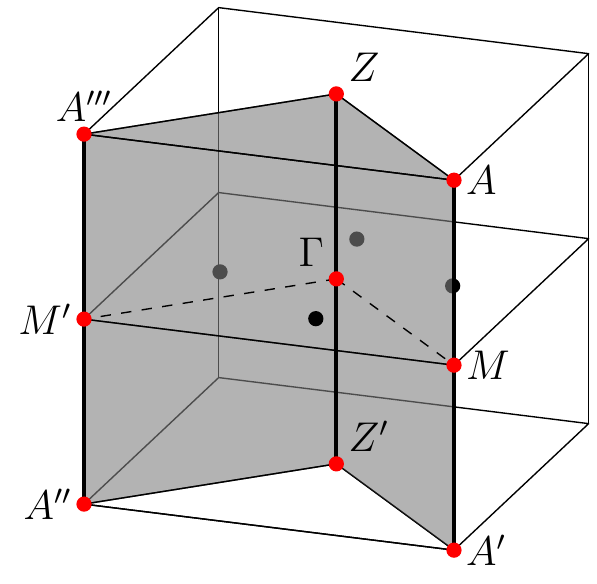}
    \caption{The full BZ with the Weyl points labeled in black dots, and the rotoinversion points labeled in red dots. Due to the $R_{4z}$ symmetry and the Weyl points, the surface $ZAA^\prime Z^\prime$ carry a Chern number of $1/2$. }
    \label{fig:quadrants}
\end{figure}

{At momentum points invariant under $\ms R_{4z}$, the Bloch states can be labeled by its eigenvalues. Focusing on the $\Gamma=(0,0,0)$ point, using the fact that $f_2(\bm k)$ is odd, this requires that (assuming $f_{1,3}(0)\neq 0$, without loss of generality) up to a common $U(1)$ phase,
\be
\ms{R}_{4z} \propto \exp\[i\theta \(\hat f_1(0)\sigma_x + \hat f_3(0)\sigma_z\) \],
\ee 
where we defined $\hat f_{1,3}\equiv f_{1,3}/\sqrt{f_1^2+f_3^2}$. Further, consistency with the $f_2(\bm k)\sigma_y$  term limits us to $\theta=0$ (for which $f_2(\bm k)$ is even under $\ms{R}_{4z}$) or $\theta=\pi/2$ (for which $f_2(\bm k)$ is odd under $\ms{R}_{4z}$).} 

{One can exclude the possibility of a trivial $\ms{R}_{4z}$ with $\theta=0$ using restrictions placed by the four Weyl points. To this end, one can divide the Brillouin zone into four quadrants related by $\ms R_{4z}$, shown in Fig.~\ref{fig:quadrants}, each of which encloses one Weyl point. Due to time-reversal symmetry, the $A A^\prime A^{\prime \prime} A^{\prime \prime \prime}$ surface does not have a Chern number, and thus for a quadrant enclosing one Weyl point, the two ``dividers" ($A A^{\prime} Z^\prime Z$ and $A^{\prime \prime \prime} A^{\prime \prime} Z^\prime Z$  ) that are related by $\ms{R}_{4z}$ each contributes a Berry flux $C=\pi$. Via the Stokes theorem, this Berry flux is the difference of the Berry phases along the vertical paths  $A A^\prime$ (or $A^{\prime \prime} A^{\prime \prime \prime}$) and $Z Z^\prime$. Each of the two paths are symmetric under $\ms{R}_{4z}$, which acts as inversion $k_z\to-k_z$, and the Berry phases can be determined by the inversion (played by $\ms{R}_{4z}$) eigenvalues at the high-symmetry points. Using known results from inversion-symmetric topological insulators,~\cite{fu-kane-2007}, in terms  the $\ms{R}_{4z}$ invariant points $\{\Gamma = (0,0,0),\  M = (\pi,\pi,0),\ Z = (0,0,\pi),\  A = (\pi,\pi,\pi)\}$, the existence of four $\ms{R}_{4z}$ symmetric Weyl point translates to
\be
\prod_{\bm k^*\in\{\Gamma, M, Z,A\},i\in \textrm{occ.}} \eta_{\bm k^*}^i=-1,
\label{eq:etas}
\ee
where $\eta_{\bm {k^*}}^i$ is the eigenvalue $\ms{R}_{4z}$ at $\bm k^*$ for the $i$-th occupied band, which takes the value of $\pm 1$ by properly choosing a common $U(1)$ phase in $\ms{R}_{4z}$. This clearly eliminates the possibility $\theta=0$, and we have
\be
\ms{R}_{4z}= \hat f_1(0)\sigma_x + \hat f_3(0)\sigma_z.
\ee

{Note that the condition \eqref{eq:etas} also eliminates the possibility of time-reversal symmetry with $\ms{T}^2=-1$ for a four-point Weyl semimetal with $\ms R_{4z}$. If $\ms{T}^2=-1$, at the above $\ms{R}_{4z}$-invariant momenta every band would be doubly degenerate via the Kramers theorem. For every given band in each of the two 1d subsystems, $(Z Z^\prime)$ and $(A A^\prime)$, since $\ms R_{4z}$ acts as spatial inversion, its eigenvalues $\eta$ at high-symmetry points can again be chosen to be $\pm 1$. Time-reversal operation either preserves the sign of $\eta$ or flips it. But since time-reversal symmetry is local and each 1d band can be represented by a 1d Wannier state, this action must be independent of $k_z=0$ or $k_z=\pi$. Therefore, time-reversal partners from every band give the same contribution to the product on the left hand side of \eqref{eq:etas}, thus incompatible with the criterion that the product for all states is $-1$.}

{As a concrete example, a lattice model with $\ms{R}_{4z}$ and $\ms T$ is given by
\begin{align}
f_1(\bm k)=&\gamma + \cos(k_z) + \cos(k_x),\nonumber\\
f_3(\bm k)=&\gamma + \cos(k_z) + \cos(k_y),\nonumber\\
f_2(\bm k)=&\sin(k_z).
\end{align}
As can be easily checked, such a model has four Weyl nodes for $-2<\gamma<0$. In this case 
\be
\ms R_{4z} = (\sigma_x+\sigma_z)/\sqrt{2},
\ee and indeed the condition Eq.~\eqref{eq:etas} is satisfied.
}

For later use we note that there are two additional composite symmetries, $\ms C_{2z} \equiv \ms R^2_{4z}$, and $\ms C_{2z} \ms T$ {{which generate subgroups of the full symmetry group generated by $\ms{R}_{4z}$ and $\ms{T}$.}} The symmetries act as
\begin{align}
    \ms C_{2z} =&\; -\mathbb{1}, 
    \qquad 
    \ms C_{2z} \ms T = -\mc K. 
\end{align}
In Sec.~\ref{sec:botsc} we will relax the $\ms R_{4z}$ symmetry and only impose $\ms C_2$. From the action of the $\ms C_{2z} \ms T$ on the Hamiltonian it can be seen that, 
\begin{gather}
    f_{1,3}(k_x, k_y, -k_z) = f_{1,3}(k_x, k_y, k_z) \nonumber \\ 
    f_2(k_x, k_y, -k_z) = -f_2(k_x, k_y, k_z). \label{eq:C2zT_constrains}
\end{gather}
The second line implies that the Weyl points are all located at either $k_z=0$ or $\pi$, and are therefore also related by $\ms{C}_{4z}$. For concreteness, we take the $4$ Weyl points to exist on the $k_z = 0$ plane with positions $\pm\bm K$ and $\pm\bm K^\prime$ such that $\bm K^\prime = \ms R_{4z} \bm K$. We further focus on the low-energy fermions near the Fermi surfaces by expanding the Hamiltonian near the Weyl points, 
\begin{align}
    h_{I}(\delta \bm k) \equiv \mc H_n(I + \delta \bm k) = \delta  k_i \phi_I^{ij} \sigma_j - \mu, 
\end{align}
where $I \in \{\pm \bm K, \pm \bm K^\prime\}$ is the set of Weyl-point, and $\phi_I^{ij} = \partial_{k_i}{f^j(\bm k)}\big|_{\bm k = I}$. The chirality of the Weyl points is given by $\sgn [\det \phi^{ij}_I]$. For later convenience, we define, 
\begin{align}
    \epsilon_I(\delta \bm k) =&\; \sqrt{\delta k_i [\phi_I \phi^{T}_I]^{ij} \delta k_j}  \\ 
    \xi_I(\delta \bm k) =&\; \epsilon_I(\delta \bm k) - \mu \\ 
    \hat n^{i}_I(\widehat{\delta \bm k}) =&\;\frac{\delta  k_j \phi_I^{ji}}{\epsilon_I(\delta \bm k)}.
\end{align}

\subsection{Pairing instability}
\begin{figure}[t]
    \centering
    \includegraphics[scale=0.9]{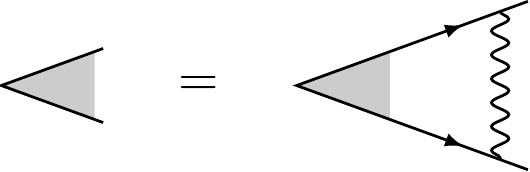}
    \caption{The linearized gap equation for the Cooper pairing vertex. See Eq.~\eqref{eq:gap}.}
    \label{FIG:cooper_instability}
\end{figure}
For a finite proper chemical potential, each of the Weyl points will be surrounded with an ellipsoidal Fermi surface (FS). Let us consider the Cooper instabilities of such a WSM model in the presence of a finite-range attractive density-density interaction. The interaction is given by 
\begin{align*}
    H_{\text{int}} = - \int d\bm k d\bm k' d\bm q\,\psi_{\bm k,\alpha}^\dagger  \psi_{\bm k+\bm q,\alpha} V(\bm q) \psi_{\bm k'+\bm q,\beta}^\dagger  \psi_{\bm k',\beta},
\end{align*}
where $\alpha,\beta$ denotes pseudospin indices, and  the attractive potential depends on momentum transfer $\bm q$. The range of the interaction is characterized by the inverse width of the peak of  $V(\bm q)$ around $\bm q=0$.
For our purposes, the relevant momentum transfer are those that connect electrons on the Fermi surfaces. In the limit where $\mu$ is small, it is a good approximation to take the interaction to only depends on which of the Fermi surfaces the two electrons belong to. We define, $V_{II^\prime} \equiv V(I - I^\prime)$ as the interaction between an electron on the $S_I$ Fermi surface and another on the $S_{I^\prime}$ Fermi surface. Due to the $\ms R_{4z}$ symmetry, we have 
 \begin{align}
    V_{II^\prime} = \mqty(V_0 && V_1 && V_2 && V_1 \\ 
    V_1 && V_0 && V_1 && V_2 \\ 
    V_2 && V_1 && V_0 && V_1 \\ 
    V_1 && V_2 && V_1 && V_0)_{II'}
    \label{eq:VII'}
 \end{align}
The pairing Hamiltonian is written as,
\begin{subequations}
\be
H_{\Delta} =  \int d \bm k \psi_{\bm k}^\dagger  \Delta(\bm k) [\psi_{-\bm k}^\dagger ]^T+ \mathrm{H.c.}.
\ee
Analogous to spin-singlet and triplet pairing, one can conveniently express $\Delta(\bm k)$ via
\begin{align}
    \Delta(\bm k) =  [\Delta^0(\bm k) + \bm d(\bm k)\cdot \bm \sigma]i\sigma_y,
\end{align}
\end{subequations}
although here due to the lack of $SU(2)$ symmetry in the band space, the four components are in general mixed. In the weak coupling limit, the linearized gap equation is given by
\begin{align}
    \Delta(\bm k) = 
    {T_c} \int_{\underline{k}'} V(\bm k - \bm k') G(\underline{k}') \Delta(\bm k') G^T(-\underline{k}'),
    \label{eq:gap}
\end{align}
where $\underline{k}'\equiv (\bm{k}',\omega_m)$ and $\int_{\underline{k}'}$ is a shorthand for the integral over momenta $\bm{k}'$ and the Matsubara sum over frequencies $\omega_m=(2m+1)\pi T$. The Green function $G(\underline{k})\equiv G(\bm k,\omega_m) =- [i\omega_m - \mathcal{H}_n(\bm k)]^{-1}$.
Using time reversal symmetry we have, 
\begin{align}
    G^{T}(-\bm k, \omega_m) = G(\bm k, \omega_m).
\end{align}
which can be used to simplify the form of the gap equation. Further,  the Green's functions can be approximated by projecting onto the low-energy electrons making up the FS's:
\begin{align}
    G_I(\delta \bm k,\omega_m)= -\frac{P_I(\widehat{\delta \bm k})}{i\omega_m-\xi_{I}(\delta \bm k)},
\end{align}
where $P_I(\delta \bm k)$ is the projection operator onto the states near the Fermi surface, 
\begin{align}
    P_I(\widehat{\delta \bm k}) = \frac{1}{2}\(\mathbb{1} + \hat{\bm n}_I(\widehat{\delta \bm k})\cdot \bm \sigma\).
\end{align}
The momentum integral can be restricted to the vicinity of the four Weyl FS's, {on which we assume} $\Delta(\bm k)$ takes constant values, and we have 
\begin{align}\label{eq:SCSimplified}
    \Delta_I= T_c\sum_{\omega_m, I^\prime} \int d\delta \bm k \ 
    V_{I I^\prime}P_{I^\prime}(\widehat{\delta \bm k})\frac{ \Tr[P_{I^\prime}(\widehat{\delta \bm k}) \Delta_{I^\prime}]}{\omega^2_m+\xi^2_{I^\prime}(\delta \bm k)},
\end{align}
where we define $\Delta_I = \Delta(I)$. {Thus, the pairing gap equation in general reduces to an eigenvalue problem for a 16 component vector  (four components $(\Delta^0, \bm d)$ for each Weyl point $I$), and strongest pairing tendency corresponds to the channel with the largest eigenvalue $T_c$. Using the fact that $\hat{\bm n}_I(\delta \bm k)$ is odd in $\delta \bm k$, we notice that \emph{independent} of the details of the band structure, $\Delta_{I} = d^y_{I} \mathbb{1}$, i.e., the ``triplet channel" with $\bm d=d^y\hat y$ is always an eigenmode of Eq.~\eqref{eq:SCSimplified}.}

{In fact, as we prove in Appendix \ref{app:cooper_leading_instability}, as long as the range of the interaction is sufficiently longer than the lattice constant (such that $V_0$ is the dominant component in Eq.~\eqref{eq:VII'}), the \emph{leading} instability of the system which gaps out all the Fermi surfaces is of the $\Delta_I = d^y_I \mathbb{1}$ type. Compatible with the Fermi statistics $\Delta_{-I} = -\Delta_{I}$, we found that such a state is an irreducible representation of $\ms R_{4z}$ that transform as
\begin{align}\label{eq:R4z_pairing}
    \ms R_{4z} \Delta_{I} \ms R_{4z}^T = \pm i \Delta_{\ms R_{4z} I},
\end{align}
and the choice of $\pm i$ spontaneously breaks $\ms{T}$. This is analogous to the $p_x+ip_y$ pairing order for inversion symmetric systems. As we show in Appendix \ref{app:cooper_leading_instability}, the superconducting critical temperature is given by
\begin{align}
    T_c = \Lambda \exp[-\frac{2}{(V_0 - V_2)N(0)}],
\end{align}
where $\Lambda$ is an upper cutoff either from the band structure or from the interaction.} {We write the pairing gap as
\begin{gather}
    \Delta(\bm k) = \(\Delta_1(\bm k) + i \Delta_2(\bm k)\) \mathbb{1},
\end{gather}
and the BdG Hamiltonian as
\begin{align}\label{eq:general_bgd_ham}
    \mc H(\bm k) = \bm f(\bm k) \cdot \bm \sigma \tau_z - \mu \tau_z + \Delta_1 (\bm k) \tau_x + \Delta_2(\bm k) \tau_y,
\end{align}
where the real gap functions are odd in $\bm k$:
\be
\Delta_{1,2}(-\bm k) = -\Delta_{1,2}(-\bm k),
\ee and  $\tau_i$ are the Pauli matrices in the Nambu space. The rotoinversion symmetry for the BdG Hamiltonian that satisfies Eq.~\eqref{eq:R4z_pairing} is given by
\begin{align}
    \ms R_{4z} =\(\hat f_1(0)\sigma_x + \hat f_3(0)\sigma_z\) e^{-i\frac{\pi}{4}\tau_z}.
     \label{eq:bdg_R4z}
\end{align}
The BdG Hamiltonian as always has a built-in particle-hole symmetry $\ms P = \tau_x \mc K$. \ }

\section{Higher-Order Topological superconductor with rotoinversion symmetry \texorpdfstring{$\ms{R}_{4z}$}{}}\label{sec:roto_inv_model}

In previous works \cite{Schindlereaat-2018, Yuxuan-2018, Apoorv_2020}, $\ms{C}_{2n}\ms T$ (with $n>1$) symmetric HOTIs and HOTSCs have been studied and their second order topology has been analyzed in some detail. In such systems, one often finds that when defined on a $\ms{C}_{2n}$ symmetric spatial geometry, the model support gapless chiral  modes 
along {hinges that are related by $\ms{C}_{2n}$ symmetry.} These chiral modes would intersect at points on the surface that are $\ms C_{2n}$ invariant. This point of intersection is protected by the $\ms C_{2n} \ms T$ symmetry. The present situation is slightly different. Since there are no fixed points on the surface under the rotoinversion action. The symmetry does not necessitate {any particular spatial position to host gapless modes.} However we still find a gapless chiral mode along a rotoinversion symmetric locus on the surface that is protected by the rotoinversion symmetry. This situation is somewhat similar to the case of inversion symmetric models with second order topology \cite{Eslam-2018}. 

{We now analyze the higher-order topology of the Weyl superconductor in Eq.~\eqref{eq:general_bgd_ham}. We first numerically solve for the spectrum of a concrete tight-binding model with open boundary conditions and demonstrate the existence of chiral hinge modes. Next, by investigating the irreducible represention of the little groups of $\ms R_{4z}$ at high symmetry points, we show that the system does not have a Wannier representation and is in a topological (obstructed) phase. Finally in this section we directly associate the nontrivial topology with the hinges by treating the hinges of a finite sample as defects of a space-filling system. The gapless modes hosted on the relevant hinges are naturally captured by the defect classification of topological phases.}

\subsection{Numerical Calculations of the Majorana Hinge Modes}
\begin{figure*}[t]
    \begin{minipage}[t]{0.49\textwidth}
        \centering
        \subfloat[][]{\includegraphics[scale=0.65,trim=0 0 0 20, clip]{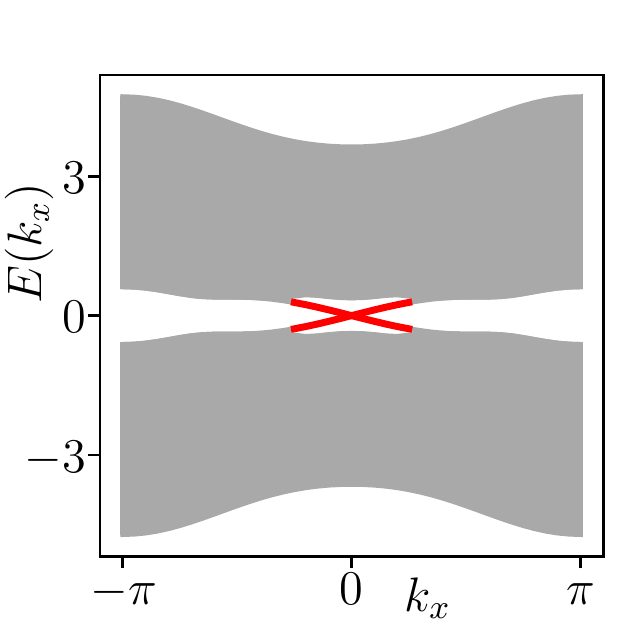}}
        \subfloat[][]{\includegraphics[scale=0.65,trim=0 0 0 20, clip]{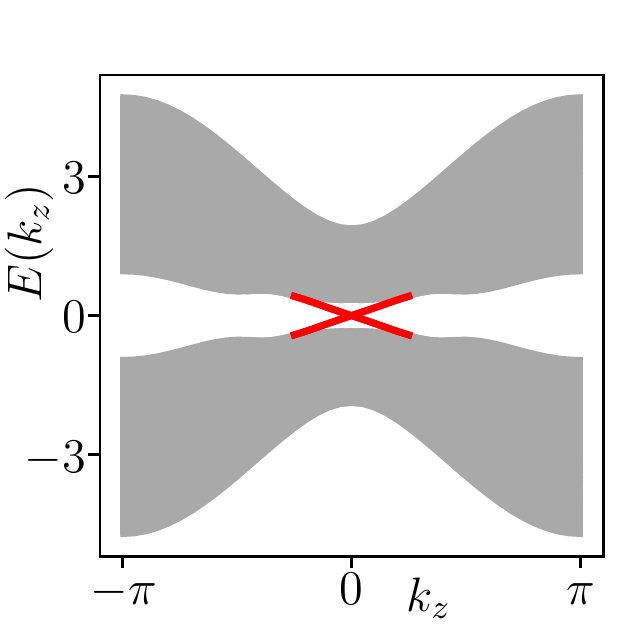}}

        \subfloat[][]{\includegraphics[scale=0.65,trim=0 0 0 5, clip]{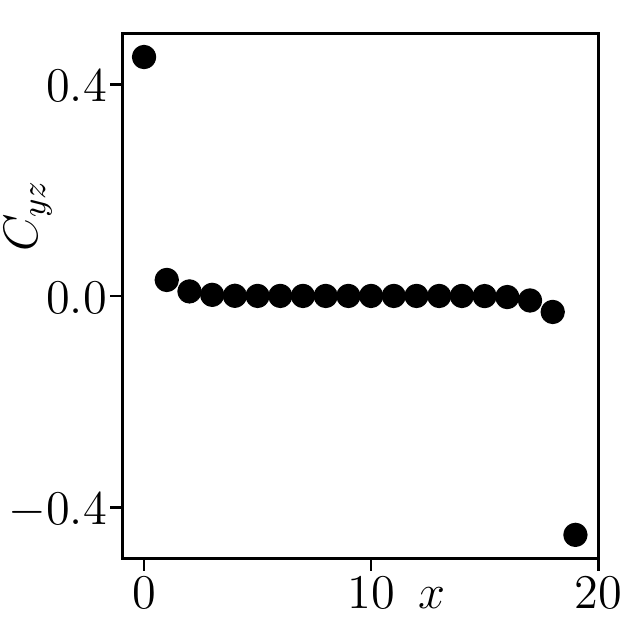}}
        \subfloat[][]{\includegraphics[scale=0.65,trim=0 0 0 5, clip]{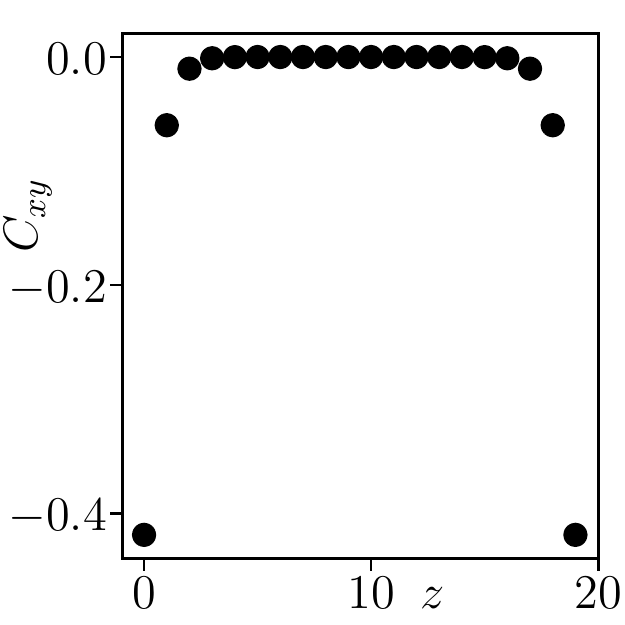}} 
        \caption{Panels (a-b) show the energy the energy spectrum with periodic boundary conditions in one direction and open boundaries in the other two. We only show $E(k_x)$, and $E(k_z)$ since $E(k_y)$ is related by $R_{4z}$ symmetry. The layer resolved Chern number is calculated in (c-d) for a disk geometry with periodic boundary conditions in two directions and open boundary condition in the third direction. We only show $C_{yz}$, and $C_{xy}$ since $C_{xz}$ is related by $R_{4z}$ symmetry.  The parameters used for the plots are $\gamma = -1, \Delta = 0.4, \mu = 0.5.$ Size = $(15 \times 15)$.}
        \label{fig:majorana_chiral_modes_4fold_sym}
    \end{minipage}\hfill
    \begin{minipage}[t]{0.49\textwidth}
        \centering 
        \subfloat[][$\mu = 0.5$]{\includegraphics[]{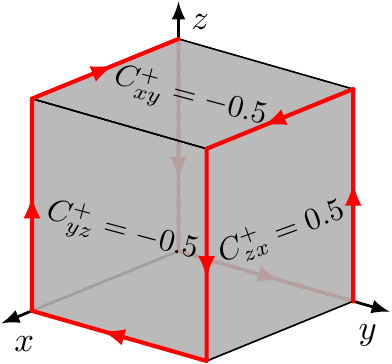}}
        \subfloat[][$\mu = 0$]{\includegraphics[]{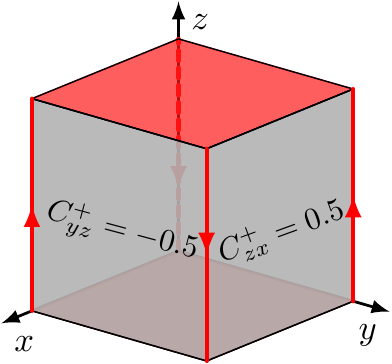}}
    
        \subfloat[][$\mu = -0.5$]{\includegraphics[]{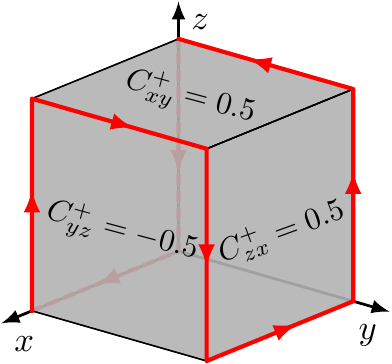}} \vspace{12pt}
        \caption{Illustration of the hinge Majorana zero modes for different vales of the chemical potential. The Majorana modes are drawn as red lines on hinges with an arrow indicating the chirality of the mode in the hinge. Hinges in black are gapped. Surfaces in gray are gapped and, those in red are gapless. When $\mu =0$ (b) the top and bottom surface are gapless. The hinge modes split in different directions depending on the sign of $\mu$ (a,c). }
        \label{Fig:hinge_modes}
    \end{minipage}
\end{figure*}
We first present numerical results on a specific tight-binding Hamiltonian which satisfies the properties discussed in the previous section. 
\begin{align}
    \mathcal{H}(\bm k) =& \[\gamma + \cos(k_z) + \cos(k_x)\] \tau_z \sigma_x + \sin(k_z) \tau_z \sigma_y \nonumber\\
    &+ \[\gamma + \cos(k_y) + \cos(k_z)\] \tau_z \sigma_z  - \mu \tau_z \nonumber\\
    &+ \Delta \sin(k_x) \tau_x  + \Delta \sin(k_y) \tau_y,
    \label{eq:mfWSM-BdG-4-fold-sym}
\end{align}
with $\gamma=1$. The $\ms R_{4z}$ symmetry for this model takes the following form, 
\begin{align}
    \ms R_{4z} = \frac{\sigma_x + \sigma_z}{\sqrt{2}} e^{-i\frac{\pi}{4}\tau_z}.
    \label{eq:R4z_defn}
\end{align}


By taking periodic boundary condition in one direction and open boundary condition in the other two we can numerically solve for the hinge modes of the Hamiltonian in Eq.~(\ref{eq:mfWSM-BdG-4-fold-sym}) using exact diagonalization. We show the results of this calculation in Fig.~\ref{fig:majorana_chiral_modes_4fold_sym}~(a-b) for the case when $\mu>0$. Chiral modes are shown in red and we find $4$ of them propagating in the $\pm k_z$ direction, and only $2$ propagating in the $\pm k_{x,y}$ directions. Further checking of the localization of these chiral modes shows that indeed they are localized in the hinges, as illustrated in Fig.~\ref{Fig:hinge_modes}(a).

We perform the same calculation but for $\mu = 0$ and $\mu<0$. The top and the bottom surfaces are gapless for $\mu = 0$. However this gap is not protected by the $\ms R_{4z}$ symmetry, and depending on $\sgn \mu$, the top and bottom surfaces become gapped in different ways as shown in Fig.~\ref{Fig:hinge_modes}(a,c).

To better understand the topology of the system, we calculate the layer resolved Chern number on the $n$-th layer of a slab geometry defined as,
\begin{align}
    C_{ij}(n) = \frac{\Imm}{\pi}  \int_{\bm k_{||}} \Tr \[\mc P(\bm k_{||})  \partial_{k_i} \mc P(\bm k_{||})\mc P_n \partial_{k_j} \mc P(\bm k_{||})\] 
\end{align}
where $\bm k_{||} = (k_i, k_j)$, are the components of the momentum parallel to the $n$-th layer, $\mc P(\bm k_{||})$ is the projection operator onto the occupied bands in the slab geometry, and $\mc P_n$ is the projection operator on the $n$-th layer. The result of this calculation for slabs parallel to the $yz$, and $xy$ planes are shown in Fig.~\ref{fig:majorana_chiral_modes_4fold_sym}(c,d). A \emph{surface Chern number} can be defined as, 
\begin{align}\label{eq:surface_chern}
    C^{\pm}_{ij} = \sum_{n \in S^{\pm}} C_{ij}(n),  
\end{align}
where $S^{\pm}$ is the set of upper/lower half of the layers. The layer resolved Chern numbers vanish for the bulk layers, hence we interpret $C^{\pm }_{ij}$ as a surface quantity.
Restrictions imposed by $\ms R_{4z}$ imply, 
\begin{align}
    C^{+}_{xy} = C^{-}_{xy}, && C^{+}_{yz} = - C^{+}_{zx}, && C^{+}_{zx} = C^{-}_{yz}.
\end{align}
Combining the above restrictions with the requirement that a chiral Majorana modes arises on the interface where this surface Chern number changes by $\pm 1$, we get that all $C_{ij}^{\pm}$ are fixed to be either $\pm 0.5$.

For the quasi-2D slab geometry with open boundary conditions in one direction, the total Chern number can be obtained by summing over all layers and are integers as expected. In the $x$ and $y$-directions the total Chern number is zero, the total Chern number with open boundary conditions in the $z$-direction is $-\sgn \mu$, for a small $\mu$. This is despite the fact that the bulk (when periodic boundary conditions are taken in all directions) has zero Chern number on all planes in the Brillouin zone. Projecting the Majorana Chiral modes in Fig. \ref{fig:majorana_chiral_modes_4fold_sym}(a,c), onto the $xy$-plane, one ends up with a Chiral Majorana mode circling the edges of the sample in a clockwise, or anti-clockwise  fashion, consistent with the positive, or negative value of $\mu$ used in this calculation.

Next, our goal is to show that the existence of the higher-order topological phase only depends on the low energy properties of the model in Eq.~(\ref{eq:general_bgd_ham}) and not on the specifics of the tight-binding model discussed here.

\subsection{Wannier obstruction}
The pairing terms in the BdG Hamiltonian in Eq.~(\ref{eq:general_bgd_ham}) break time-reversal symmetry, thus with only $\hat{\ms P}^2 = 1$ the system is in the AZ symmetry class D. Since $3$D class D systems do not support non-trivial band topology, there is no obstruction to having a well localized Wannier representation. The meaning of the Wannier representation for BdG Hamiltonian has been previously studied~\cite{Geier_2020, Skurativska_2020, Schindler_2020}. We therefore ask whether there exists a Wannier representation that respects the $\ms R_{4z}$ symmetry as well. We check this using a symmetry indicator approach. If such a Wannier representation exists, the centers of the Wannier functions should reproduce the eigenvalues of the symmetry operators at the high-symmetry points on the Brillouin zone. As mentioned before, there are four points in the Brillouin zone that are invariant under $\ms R_{4z}$, $\{\Gamma = (0,0,0),\  M = (\pi,\pi,0),\ Z = (0,0,\pi),\  A = (\pi,\pi,\pi)\}$. All of the $\ms R_{4z}$ invariant points are also time-reversal invariant, and thus the pairing terms vanish and the Hamiltonian take the following form
\begin{align}
     H(\bm k^*) = \bm f(\bm k^*) \cdot \bm \sigma \tau_z 
\end{align}
where $\bm k^* \in \{\Gamma, M, Z, A\}$.

The eigenvalues of the rotoinversion symmetry operator for the occupied bands of the Hamiltonian given in Eq.~(\ref{eq:general_bgd_ham}) is shown in Fig \ref{fig:rotoinversion_eigenvalues}.
We notice that the symmetry operators eigenvalues are completely determined by $\eta_{\Gamma}$, $\eta_{M}$, $\eta_{Z}$, and $\eta_{A}$. On the other hand, the pairing terms ensure that the system is completely gapped, and modifies the form of the $\ms R_{4z}$ operator to that in Eq.~\eqref{eq:bdg_R4z}. After a straightforward enumeration of all the possible Wannier centers and the resulting $\ms R_{4z}$ eigenvalues we find

\begin{align} \label{eq:obstruction_condition}
    &\eta_\Gamma \eta_M \eta_Z \eta_A = 
    \begin{cases}
        -1, &  \text{obstructed} \\
        1, & \text{not obstructed}.
    \end{cases}
\end{align} 

\begin{figure}[t]
    \centering
    \includegraphics[scale=1.2]{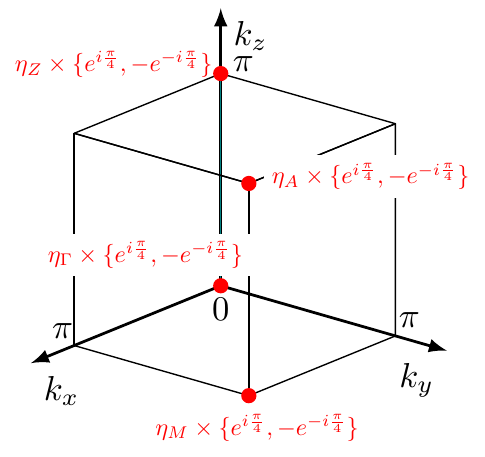}
    \caption{Occupied states symmetry operators eigenvalues at the high symmetry points in the Brillouin zone. In red are the eigenvalues of $\ms R_{4z}$ operator at the rotoinverison invariant points, and in teal are the eigenvalues of $\ms C_{2z}$ on the twofold rotation invariant lines.}
    \label{fig:rotoinversion_eigenvalues}
\end{figure}

The obstruction in the system can be understood as follows: consider a \emph{hybrid} Wannier representation of the system that is localized in the $x$ and $y$-directions but not in the $z$-direction, $\ket{\nu^i(R_x,R_y,k_z)}$, $i \in \{1,2\}$. At $k_z = 0, \pi$ the $\ms R_{4z}$ symmetry reduces to a fourfold rotation symmetry, $\ms R_{4z}\ket{\nu^i(R_x,R_y,k_z = 0,\pi)} =$ $ \ket{\nu^i(-R_y,R_x,k_z = 0,\pi)}$. Similar $2$D systems under the restriction of fourfold rotation symmetry are studied in~\oncite{tiwari2020chiral}. The Wannier functions for the $2$D subsystem at $k_z =0 \ (\pi)$ are either centered at $r = (0,0)$ when $\eta_{\Gamma} \eta_{M} = 1$ ($\eta_{Z} \eta_{A} = 1$), or at $r = (1/2,1/2)$ when $\eta_{\Gamma} \eta_{M} = -1$ ($\eta_{Z} \eta_{A} = -1$), where $r$ is measured relative to the unit cell center. The condition for obstruction is that only one pair, either $\eta_\Gamma,$ and $\eta_M$, or  $\eta_Z,$ and $\eta_A$ have a relative minus sign, but not both. As we discuss below, the existence of the  Weyl points in the $k_z = 0$ plane but not in the $k_z = \pi$ plane, ensures a relative minus sign between $\eta_\Gamma,$ and $\eta_M$. Thus, in this hybrid Wannier reprsentation, the Wannier centers are centered at $r = (1/2,1/2)$ at $k_z = 0$, and as we increase $k_z$ the Wannier centers drift and reach $r=(0,0)$ at $k_z = \pi$. This kind of \emph{Wannier spectral flow} indicates that the system cannot be further localized in the $z$-direction.

{Interestingly, we note that the condition for Wannier obstruction is precisely the one in Eq.~\eqref{eq:etas} we found for the existence of four Weyl points related by $\ms{R}_{4z}$. Therefore, generally we have proven that an $\ms R_{4z}$ Weyl semimetal with four Weyl nodes with attractive interaction naturally host a higher-order topological superconducting phase. This is the main result of our work. }

\subsection{Gapless hinge modes from defect classification}
In this section we analyze the topology of the model in Eq.~(\ref{eq:general_bgd_ham}), from its defect classification. We treat the appearance of stable gapless states at codimension-1 or higher as a diagnostic of non-trivial bulk topology. In particular we are interested in the appearance of gapless chiral hinge modes on $\ms{R}_{4z}$ symmetric hinges on the surface of an open geometry. To this end, consider placing the model on an open geometry that preserves the rotoinversion symmetry. Outside the sample exists a perfectly featureless atomic insulator that also preserves the spatial symmetry. As the outside region is featureless, the four Weyl-points must annihilate somewhere along the surface of the sample. Since we insist on preserving the rotoinversion symmetry, the Weyl-points are forced to annihilate at one of the four $\ms{R}_{4z}$-symmetric points $\bm k^*= \Gamma,M,Z$ or $A$. 

In any of these cases, the low energy physics is described by keeping only the leading order terms in a small momentum expansion $\delta \bm k$ from the rotoinversion invariant point. We define,
\begin{align}
    f_{1,3}(\bm k^*+ \delta \bm k) =&\; \ms m_{1,2}, \nonumber \\ 
    f_2(\bm k^* + \delta \bm k) =&\;  v_z \delta k_z \nonumber  \\ 
    \Delta_{1,2}(\bm k^* + \delta \bm k) =&\;  v^{1,2}_x \delta k_x + v^{1,2}_y \delta k_y,
\end{align}
where we used the evenness of $f_{1,3}(\bm k)$ (Eq.~\eqref{eq:TRS-constrains}) and the fact that $f_{2}(\bm k)$ is zero over the entire $\bm k_z = 0, \pi$ planes (Eq.~\eqref{eq:C2zT_constrains}) from which it follows that it has no linear terms in $ k_x$ and $ k_y$ on these planes. Furthermore, from the odd parity nature of the pairing, and upon applying Eq.~\eqref{eq:R4z_pairing} twice we obtain that $\Delta_{1,2}(\bm k)$ are even under $k_z \rightarrow - k_z$, and thus have no linear terms in $k_z$. 

From the action of the rotoinversion symmetry we see that $\ms R_{4z}: \bm v^2 \rightarrow \bm v^1$, where $\bm v^{1,2} = (v^{1,2}_x, v_y^{1,2},0)$, meaning $v^2_x = v^1_y = v_y$, and $v^2_y = -v^1_x = -v_x$. The low energy continuum Weyl model in the vicinity of the rotoinversion invariant point takes the form 
\begin{align}
     H(\delta \bm k)=&\; v_{xy} \left(\delta k_x \gamma^{1}+\delta k_y \gamma^{2} \right) + v_z \delta k_z \gamma^{3} \nonumber \\ 
    &\;  + \ms{m}_{1}\gamma^{4} + \ms{m}_{2}\gamma^{5} - \mu \gamma^{12},
\label{Eq:low_energy_Ham}
\end{align} 
where for convenience we define $v_{xy} = \sqrt{v_x^2 + v_y^2}$, and 
\begin{gather}
    \gamma^1 = \frac{1}{v_{xy}}(v_x \tau_x + v_y \tau_y),\quad  \gamma^2 = \frac{1}{v_{xy}}(v_y \tau_x - v_x\tau_y), \nonumber \\ 
    \gamma^3 = \sigma_y \tau_z, \quad \gamma^{4,5} = \sigma_{x,z}\tau_z, \quad \gamma^{1,2} = i\gamma^1 \gamma^2
\end{gather}
In the bulk, the mass vector $\bm{\ms m} = (\ms m_1, 0, \ms m_2) $ is constrained such that  $ \bm{\ms m} = \pm \ms m(\hat{f}_1(0),0 , \hat{f}_3(0))$, with $\ms{m}^{2}=\ms{m}_{1}^{2}+\ms{m}_2^{2}$. However, it may vary as one approaches the surface. If $\bm{\ms m}(\bm r)$ represents the mass domain wall close to the surface, then $\bm{\ms m}(\bm r)$, and $\bm{\ms m}(\ms R_{4z} \bm r)$ are related by a reflection about the $(\hat{f}_1(0),0 , \hat{f}_3(0))$ direction. 

Below we present two complementary approaches to study the existence of hinge modes. The first approach is based on the notion of dimensional reduction/adiabatic pumping while the second approach makes use of a classification of line defects in BdG superconductors. 
\subsubsection{Via adiabatic pumping}
\medskip In this section we show that the 3D class $\ms{D}$ hinge superconductor in Eq.~\eqref{Eq:low_energy_Ham} dimensionally reduces to a class $\ms{BDI}$ second-order superconductor in 2D which was studied in Ref.\cite{tiwari2020chiral}. The roto-inversion $\ms{R}_{4z}$ reduces to a fourfold rotation $\ms{C}_4^{z}$ in the $x$-$y$ plane. In order to make this dimensional transmutation precise, we write the low energy Hamiltonian \eqref{Eq:low_energy_Ham} in the following suggestive way by replacing $\delta k_{z}\to -i\partial/\partial z$
\begin{align}
 H(\delta {k}_x, \delta {k}_y, z)= H_{2\text{D}}(\delta k_{x},\delta k_{y}) + i v_z \gamma^{3}\frac{\partial}{\partial z}.
\end{align} 
We first consider setting the chemical potential $\mu=0$. With $\mu=0$, note that the Hamiltonian $\mathcal H_{2\text{D}}$ describes a class $\ms{BDI}$ superconductor. This is due to the fact that since $\left\{\gamma^{3},\bm H_{2\text{D}}(\bs{k})\right\}=0$, $\gamma^{3}$, effectively implements a chiral symmetry for the 2D model. Moreover it was shown in Ref.\cite{tiwari2020chiral} that this model describes a $\ms{BDI}$ second-order superconductor that supports Majorana zero-modes at the corners of a $\ms{C}_{4}^{z}$ symmetric spatial geometry. The states localized at each corner can be indexed by an integer  $\ms{N}_{\ms{w}}\in \mathbb Z_{\text{odd}}$ which corresponds to the difference in the number of zero-energy eigenstates with positive and negative chirality. Here we show that each such mode contributes to a chiral gapless mode on the hinge of the 3D model. Consider the ansatz of the form $|\Psi(k_{x},k_{y},z,t)\rangle =\phi(z,t)|\varphi(k_{x},k_{y})\rangle$ where $|\varphi(k_{x},k_{y})\rangle$ is a zero-mode of the 2D model with chirality $+1$, i.e $\bm H_{2\text{D}}(k_x,k_y)|\varphi(k_x,k_y)\rangle=0$ and $\Gamma^{3}|\varphi(k_x,k_y)\rangle=|\varphi(k_x,k_y)\rangle$.  Then solving the Schrodinger equation gives $\phi(z,t)=\phi(z+t)$. Similarly one obtains $\ms{N}_{\ms{w}}$ chiral Majorana modes with opposite chirality on adjacent corners.  

The discussion above survives if we turn on a  small but finite chemical potential. Indeed it was shown in Ref.~\cite{tiwari2020chiral}, that that the corresponding Hamiltonian $\mathcal H_{\text{2D}}$ has majorana zero modes present at the corners of a $\ms{C}_{4z}$ symmetric spatial geometry. The topological invariant associated to these zero modes is the mod 2 reduction of the winding number $\ms{N}_{\ms{w}}$ \cite{Shiozaki_index}. The chirality of the hinge mode remains unchanged as compared with $\mu=0$ case since it cannot change without a gap opening. In the next section we describe an alternate approach that provides a diagnostic for the higher-order topology based on the defect classification.

\subsubsection{Defect invariant:  Second Chern number}
\noindent Let us formulate \eqref{Eq:low_energy_Ham} as a continuum Euclidean time Dirac action 
\begin{align}
S=&\; \int\mathrm{d}^{3}x\mathrm{d}\tau \Psi^{\dagger}\left[\partial_{\tau} + i\sum_{i=1}^{3}\gamma^{i}\partial_{i} + \ms m_1 \gamma^4 + \ms m_2 \gamma^5 \right]\Psi,
\end{align}
defined on an open spatial geometry $M$ embedded in a trivial insulator. We absorb the velocities, $v_z$, and $v_{xy}$ through an appropriate rescaling of the coordinates. Such process does not affect the topology of the system.

Comparing $\eta_{\bm k^*}$ in the bulk and outside $M$ i.e in the region that hosts the trivial model, they differ by a minus sign. It is known that line defects in class A and class D insulators and superconductors are integer classified and host chiral Dirac and Majorana modes respectively. Moreover the integer invariant corresponding to a model containing a non-trivial defect is captured by the second Chern number evaluated on the hybrid four-dimensional space $\mathrm{BZ}\times S_{\gamma}^{1}$ where $\mathrm{BZ}$ is the 3D Brillouin zone and $S^{1}_{\gamma}$ is a real-space loop (homotopic to a circle) that links with the defect under consideration.  

Such a defect invariant can directly be applied to the study of second-order topological phases in 3D by simply considering the hinge as a defect. The role of the spatial symmetries then is to ensure the stability of the defect at particular high symmetry loci on the surface of the topological phase. We consider $S^{1}_{\gamma}$ to be a path linking with a chosen hinge. For convenience we choose a path that intersects the boundary of the spatial geometry $M$ at two $\ms{R}_{4z}$ related points. Let $\theta$ be an angular variable parameterizing the path $S^{1}_{\gamma}$. The invariant associated with the hinge, denoted as $\ms{N}_{\text{Hinge}}$ takes the form 
\begin{align}
\ms{N}_{\text{Hinge}}=&\; \frac{1}{8\pi^{2}}\int_{\mathrm{BZ}\times S^{1}_{\gamma}}\Tr\left[\mathcal F \wedge \mathcal F\right] \nonumber \\
=&\; \frac{1}{8\pi^2}\int_{\mathrm{BZ}\times S^{1}_{\gamma}}\Tr\left[\mathcal P \mathrm{d}\mathcal P \wedge \mathrm{d}\mathcal P \wedge \mathrm{d}\mathcal P \wedge \mathrm{d}\mathcal P\right],
\label{eq_inv}
\end{align}
where $\mathcal P=\sum_{i=1,2}|u_{i}(\bs{k},\theta)\rangle \langle u_{i}(\bs{k},\theta)|$ is the projector onto the occupied states $|u_{i}(\bs{k},\theta)\rangle$. In order to compute the invariant we modify our model without closing the energy gap thereby leaving the topology unaltered. More precisely, we consider the Hamiltonian 
\begin{align}
    \widetilde{         H}=\sum_{i=1}^{5}h_{i}(\bs{k},\theta)\gamma^{i},
\end{align}
where 
\begin{align}
    h_{i}=
    \begin{cases}
        \frac{k_{i}-\epsilon \bs{k}^{2}}{\sqrt{\bs{k}^{2}+\ms{m}^{2}}} &\text{if $i=1,2$}\\
        \frac{k_{i}}{\sqrt{\bs{k}^{2}+\ms{m}^{2}}} &\text{if $i=3$} \\
        \frac{\ms{m}_{i-3}(\theta)}{\sqrt{\bs{k}^{2}+\ms{m}^{2}}} &\text{if $i=4,5$} 
    \end{cases}
\label{eq:coods}
\end{align}
The term $\epsilon \bs{k}^2(\gamma^{1}+\gamma^{2})$ has been added as a $\ms{R}_{4z}$ symmetric regularization that implements a one point compactification of $\mathrm{BZ}\times S^{1}_{\gamma}$ such that $f$ denotes a map from $S^{4}$ to $S^{4}$. We take $\epsilon \to 0$ at the end of the calculation.  Additionally, we choose a path $S^{1}_{\gamma}$ on which $\ms{m}^{2}=\ms{m}_{1}^{2}+\ms{m}_2^{2}$ is independent of $\theta$. The Hamiltonian $\widetilde{ H}$ has the advantage that it is normalized with a pair of degenerate eigenstates with eigenenergies $\pm 1$. The projector onto occupied states can explicitly be written as $\mathcal P=\frac{1+\bm{h}\cdot \bs{\gamma}}{2}$. Inserting this into the expression \eqref{eq_inv} one obtains 
\begin{align}
    \ms{N}_{\text{Hinge}}=&\; \frac{1}{8\pi^2}\int \epsilon^{ijklm}h_{i} \partial_{k_{x}}h_{j}\partial_{k_y}h_{k}\partial_{k_z}h_{l}\partial_{\theta}h_{m} \nonumber \\
    =&\; \frac{1}{2\pi} \int_{S^1_{\gamma}} \bm{\ms{m}}\partial_{\theta} \bm{\ms{m}},
\end{align}
therefore the topological invariant associated with a given hinge reduces to the topological winding number associated with the map $\bm{\ms{m}}:\theta \in S^{1}_{\gamma}\to S^{1}_{\ms{m}}$ where $S^{1}_{\ms{m}}$ is the circle coordinates $\text{arctan}(\ms{m_2}/\ms{m}_1)$. {Since (1) $\ms{R}_{4z}$ acts as a reflection along the $(\hat{f}_1(0),0 , \hat{f}_3(0))$ direction on the space of masses, and (2) $\bm{\ms m}$ reverses direction when moving from deep into the bulk to far outside the sample, the winding number around the loop $S^{1}_{\gamma}$ is pinned to be an odd number \cite{tiwari2020chiral}.} To conclude we have shown that the second Chern number in hybrid space $(\boldsymbol{k},\theta)$ serves as a topological invariant which may be used to diagnose the presence of chiral Majorana hinge modes. For the Hamiltonian of the form Eq.~\eqref{Eq:low_energy_Ham} it reduces to the mass winding number around $\theta$ which is enforced to be non-vanishing and odd by the spatial $\ms{R}_{4z}$ symmetry. 
 
 \section{Classification of \texorpdfstring{$\ms{R}_{4z}$}{}-symmetric higher-order superconductors}
 \label{sec:class}
 
In this section we derive the classification of $\ms{R}_{4z}$-symmetric higher-order phases. We treat the appearance of robust ingappable modes on high symmetry lines and points on the surface of a fully gapped and spatially symmetric superconductor as diagnostics of second and third order topology respectively. For the purpose of classification, it is convenient to work with ground states directly rather than with Hamiltonians \cite{Hermele_1, Hermele_2, Hermele_3, Shiozaki_1, Shiozaki_2, Thorngren_1, Thorngren_2, Thorngren_3}. A ground state of a model within a certain topological phase with a given crystalline symmetry $\ms{G}$ can be adiabatically deformed to a particular type of state known as {\emph{block state}}. A block state corresponding to a higher-order topological phase can be understood hueristically as a network of lower dimensional topological states with only internal symmetries glued together in a manner that is compatible with all spatial symmetries.  

Here we illustrate the construction for the case of $\ms{R}_{4z}$-symmetric class D superconductors. To do so, we consider a $\ms{R}_{4z}$-symmetric cell complex, illustrated in Fig.~\ref{cell_complex}.  Since we are interested in higher-order topology and therefore boundary  modes, we consider the cellulation of an open $\ms{R}_{4z}$ symmetric geometry. The cell complex consists of a network of  1-cells and 2-cells. Note that we do not consider 3-cell as (i) they do not affect the classification of higher order phases and (ii) for the present case, i.e class D, there are no topologically non-trivial phases in 3D. Moreover, we also do not consider bulk 0-cells since they do not contribute to any boundary signatures. We consider a cell complex such that each $p$-cell is either left entirely invariant or mapped to another $p$-cell under the under the action of $\ms{R}_{4z}$. Since, the $\ms{R}_{4z}$ only has a single fixed-point, and we do not consider $0$-cells, all the $p$-cells we consider transform to $\ms{R}_{4z}$ related $p$-cells under the symmetry action.  It is therefore convenient to divide up the $p$-cells into $\ms{R}_{4z}$ orbits. There are 3 bulk and 4 boundary 2-cell orbits which in Fig.~\ref{cell_complex}, we denote as $\alpha,\beta,\gamma$ and $a,\dots,d$  respectively. Likewise there are 2 bulk and 9 boundary 1-cell orbits which we denote as $\Lambda_{1,2}$ and $\ms{A},\dots, \ms{G}$ respectively.  
\begin{figure}[t] 
\centering
\includegraphics[scale=0.6]{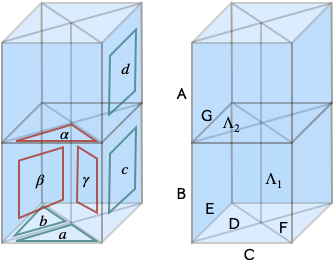}
   \caption{An illustration for a choice of cell complex for the point group $\ms{R}_{4z}$ on an open geometry. The 2-cells are illustrated in panel (a) wherein the 2-cells $a,b,c,d$ are each a representative of a certain $\ms{R}_{4z}$ orbit on the surface of the depicted geometry. Likewise $\alpha, \beta$ and $\gamma$ each label a certain $\ms{R}_{4z}$ orbit in the bulk of the geometry. Similarly, panel (b) illustrates the distinct $\ms{R}_{4z}$ orbits for the 1-cells. The representatives of the surface orbits are denoted $\ms{A},\ms{B},\dots,\ms{G}$ while the bulk orbit representatives are denoted as $\Lambda_{1}$ and $\Lambda_{2}$. }
    \label{cell_complex}
\end{figure}
 
A particular bulk state is constructed by populating a chosen orbit or more generally a collection of orbits by non-trivial topological states with the constraint that the bulk be fully gapped for the chosen network. More concretely, since class D superconductors in 1D and 2D are $\mathbb Z_{2}$ and $\mathbb Z$ classified respectively with the 1D Kitaev chain and the 2D $p\pm ip$ superconductors as generators, we may populate the bulk of the $\ms{R}_{4z}$-cellulation with states corresponding to the $p\pm ip$ and Kitaev phase on some combination of the $\alpha,\beta,\gamma$ and  $\Lambda_{1,2}$ orbits respectively. Let the state assigned to the $\alpha$-orbit have topological index $n_{\alpha}\in \mathbb Z$ and similarly for $\beta$ and $\gamma$, likewise we denote the index assigned to the 1-cells belonging to the orbits $\Lambda_{1,2}$ as $m_{1,2}$. A priori bulk states are therefore labelled by $(n_{\alpha},n_{\beta},n_{\gamma},m_1,m_2)\in \mathbb Z^{3}\times \mathbb Z_{2}^{2}$. Since each of these candidate bulk cells contribute gapless 1D modes  or zero modes on the boundaries of the cells, we must ensure that these modes can be gapped out pairwise such that one ends up with a fully gapped bulk. Notably we require $n_{\beta}+n_{\gamma}=0$ such that the central hinge ($\Lambda_{1}$) is gapped. Upon imposing this condition, the bulk is fully gapped, since (i) the gapless modes contributed by the $\gamma$ and $\beta$ orbits on the 1-cells $\Lambda_2$, cancel out pairwise upon imposing the condition $n_{\beta}+n_{\gamma}=0$ and (ii) the gapless modes contributed by the $\alpha$ orbit cancel out pairwise. Therefore the most general fully gapped bulk state is labelled as $(n_{\alpha},n_{\beta},-n_{\beta},m_{1},m_{2})\in \mathbb Z^{2}\times \mathbb Z_{2}^{2}$.
Each non-trivial bulk cell contributes a gapless mode on the boundary such that one ends up with a network of gapless currents and zero-modes on the boundary as illustrated in Fig.~\ref{bdy_config}.

\begin{figure}[t] 
\centering
\includegraphics[scale=0.6]{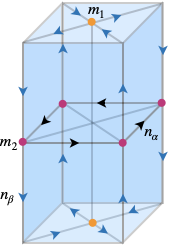}
   \caption{An illustration of a state with a fully gapped bulk a surface containing an $\ms{R}_{4z}$ symmetric configuration of chiral majorana modes and majorana zero-modes. The hinges with blue and black arrows contain $n_{\alpha}$ and $n_{\gamma}$ majorana modes while the orange and red dots denote the presence of $m_{1}$ and $m_{2}$ majorana zero modes.}
    \label{bdy_config}
\end{figure}

Next, we ask which of the above modes are truly the signature of bulk topology. To answer this question, one needs to check which modes can be annihilated or equivalently constructed from a purely surface pasting of $p\pm ip $ and Kitaev states. Firstly, it can be checked that the $m_2$ Majorana modes constributed on the surface by the presence of Kitaev state on $\Lambda_{2}$ can be trivialized by surface pasting of Kitaev chains on the orbits corresponding to the 1-cells $\ms{A}$ and $\ms{F}$. Similarly one can transform the configuration $(n_{\alpha},n_{\beta},-n_{\beta},\dots)$ to $(n_{\alpha}-n_{\beta},0,0, \dots)$ by surface pasting of $n_{\beta}$ copies of $p+
ip$ states on the $a$ and $d$ orbits. Collectively, these two operations reduce the space of non-trivial bulk states from $\mathbb Z^2\times \mathbb Z_{2}^{2}$ to $\mathbb Z\times \mathbb Z_{2}$ indexed by $(n_{\alpha}-n_{\beta},0,0,m_{1},0)$. It can be verified that the $m_1$ zero modes contributed by $\Lambda_1$ are robust, hence there exist a $\mathbb Z_{2}$ classified third order superconductor protected by point group $\ms{R}_{4z}$. Getting back to the $n_{\alpha}-n_{\beta}$ chiral majorana mode propagating around the sample on the reflection symmetric plane.  One can always change $n_{\alpha}$ to $n_{\alpha}+2n$ by pasting $n$ copies of $p\pm ip$ states on all the surface orbits $a,b,c,d$. This reduces the classification of second-order phases to $\mathbb Z_{2}$. To summarize the classification of both second and third order $\ms{R}_{4z}$ symmetric superconductors in class D is $\mathbb Z_{2}$. For second order superconductors, this is generated by the bulk state with the $\alpha$-orbit populated with $p+ip$ class D superconductors while for the third order topology, it is generated by the populating the $\Lambda_1$ orbit with Kitaev chains.    

\section{$\ms{R}_{4z}$ symmetric second-order superconductor with surface topological order}
\label{Sec:STO}
In previous sections, we showed that class D superconductors enriched by $\ms{R}_{4z}$ rotoreflection symmetry supports non-trivial second order topology. The appearance of a robust chiral majorana hinge mode on a rotoreflection symmetric line on the surface was treated as diagnostic of the second-order topology. Here we ask whether these surface modes remain robust in the presence of symmetry preserving strong interactions on the surface. We answer this question in the negative by constructing a fully gapped topologically ordered surface that preserves all the symmetries in question. We construct such a surface topological order (STO) by symmetrically introducing $\mathsf{SO}(3)_6$ non-abelian topological orders on the two $\ms{R}_{4z}$ related regions denoted $\Sigma_{1,2}$ in Fig.~\ref{fig:STO}. A similar construction for the topologically ordered surfaces of $\ms{C}_{2n}\mathcal T$-symmetric second-order topological superconductors has been previously studied in \cite{Apoorv_2020}. The $\ms{SO}(3)_6$ topological order is a `spin' or fermionic topological order \cite{Bruillard_2017} as it contains a single fermionic excitation (denoted below as $j=3$) which is local, in the sense that it braids trivially  with all other excitations/anyons in the topological order. Such a model is described by the continuum Chern-Simons action \cite{witten1989quantum, elitzur1989remarks}
 \begin{align}
 S_{I}=\frac{(-1)^{I}k}{4\pi}\int_{M_{I}}  \hspace{-10pt}\text{Tr}\left\{A\wedge dA +\frac{2}{3}A\wedge A\wedge A\right\},
 \end{align}
 where $k$ is the `level' of the Chern-Simons theory which is 6 for present purpose, $A$ is $\ms{SO}(3)$-valued gauge connection and $M_{I}=\Sigma_{I} \times S^{1}$ with $I=1,2$ labelling the two $\ms{R}_{4z}$-related regions and $S^{1}$ is the compactified time domain. 
The $\ms{SO}(3)_6$ topological order has a total of four anyons labelled $j=0,1,2,3$, with $j=3$ being a fermion 
\cite{Lukasz_2013, Wang_2017, Apoorv_2020} and $j=0$ the vacuum sector or ``trivial anyon". The $j=1,2$ anyons are semionic and anti-semionic respectively. The fusion rules among the anyons are 
\begin{align}
j\times {j}'=&\; \sum_{j''=|j-j'|}^{\text{min}(j+j', 6-j-j')}j'',
\end{align}
while the modular $\ms{S}$ and $\ms{T}$ matrices that describe the braiding and self-statistics respectively are given by
\begin{align}
\ms{T}_{j,j'}=&\; \exp\left\{2\pi i j(j+1)/8\right\}\delta_{j,j'}, \nonumber \\
\ms{S}_{j,j'}=&\; \frac{1}{2}\sin\left[\frac{(2j+1)(2j'+1)\pi}{8}\right].
\label{eq:modular_data}
\end{align}
\begin{figure}[tb] 
\centering
\includegraphics[scale=0.4]{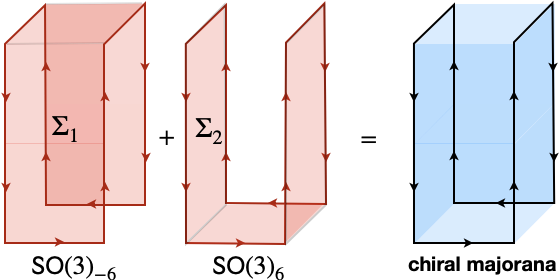}
\caption{The chiral Majorana hinge mode on the surface of an $\ms{R}_{4z}$-symmetric second-order superconductor can be gapped out by introducing a topologically-ordered surface. The figure illustrates an $\ms{R}_{4z}$-symmetric pattern of $\ms{SO}(3)_6$ topological order which furnishes a single chiral Majorana hinge mode that can gap out the hinge mode contributed from the bulk second-order superconductor.}
\label{fig:STO}
\end{figure}

Since the regions $\Sigma_{1}$ and $\Sigma_{2}$ share a common hinge as their boundary, one obtains two sets of co-propogating chiral edge modes on the hinge, one  from each of the surface topological orders. Each of these correspond to a chiral $\ms{SO}(3)_{6}$ Wess-Zumino-Witten (WZW) conformal field theory (CFT) \cite{francesco2012conformal} with chiral central charge $c_{-}=9/4$. The combined CFT on the hinge has a central charge $c_{-}=9/2$.  We denote the holomorphic current operators as $\mathcal J_{\ms{a},I}$ where $I=1,2$ again labels which topological order the mode is contributed from and $\ms{a}=1, \dots, \text{dim}(\mathfrak{so}(3))$. The current operators satisfy the operator product expansion
\begin{align}
\mathcal J_{\ms{a},I}(z)\mathcal J_{\ms{b},I}(w)\sim \frac{k\delta^{\ms{ab}}}{(z-w)^{2}} + \frac{if_{\ms{ab}}^{\ms{c}}\mathcal J_{\ms{c},I}}{z-w},
\end{align}
where $f_{\ms{abc}}$ are the structure constants of the $\mathfrak{so}(3)$ Lie-algebra. The Hamiltonian of the hinge CFT is obtained via the Sugawara construction \cite{Sugawara} and takes the form
\begin{align}
 H_0=\frac{1}{k+h^{\ms{v}}}\sum_{I,\ms{a}} \mathcal J_{\ms{a},I}\mathcal J_{\ms{a},I}.
\label{eq:sugawara}
\end{align} 
The modes of the current operators additionally satisfy the Kac-Moody algebra that acts on the states in the conformal field theory, which are thus organized into conformal towers or representations of the Kac-Moody algebra. Each representation is built on a highest weight state which is related to a conformal primary operator via the state operator map and is in one-to-one correspondence with the bulk anyons. We label the primary operators just  as the bulk anyons by a tuple $(j_{1},j_{2})$ where $j_{I}=0,1,2,3$.  One obtains conformal characters $\chi_{j_{1},j_{2}}$ by tracing over the corresponding  conformal towers $\mathcal H_{(j_1,j_2)}$ 
\begin{align}
\chi_{(j_{1},j_{2})}(\tau)=\text{Tr}_{\mathcal H_{(j_{1},j_{2})}}\left[e^{2\pi i\tau(H _0-\frac{c}{24})}\right]
\end{align}
where $H_{0}$ is the Hamiltonian in Eq.~\eqref{eq:sugawara} and $\tau$ is the modular parameter of the spacetime torus $\partial M_{I}$. The bulk topological data in Eq.~\eqref{eq:modular_data} can be recovered from the edge CFT by performing the $\ms{S}$ (i.e $\tau\to -1/\tau$) and $\ms{T}$ (i.e. $\tau\to \tau +1$) modular transformations on the conformal characters. 
Next, we deform the Hamiltonian in Eq.~\eqref{eq:sugawara} by adding terms that lead to a condensation on the hinge. Such a condensation is equivalent to adding `simple currents' to Kac-Moody algebra which furnishes a so-called {\it{extended chiral algebra}}. The simple currents that can be simulataneously condensed correspond to primary operators that are mutually local (i.e have a trivial $\ms{S}$-matrix element) and have integer spin (i.e have a trivial $\ms{T}$ matrix element). Adding simple currents to the chiral algebra further constrains the corresponding representation theory and therefore has profound physical consequences on the structure of the theory. Some of the conformal towers merge together while others are removed from the spectrum. In the present case, there are three candidate simple current operators corresponding to the primaries $(j_1,j_2)=(1,2),(2,1)$ and $(3,3)$. These primaries correspond to the only `condensable' operators as they exhaust all the integer spin operators in the theory. We denote this set as $\mathcal B$ and add the following term to the Hamiltonian in Eq.~\eqref{eq:sugawara}
\begin{align}
 H= H_0+\lambda\sum_{(j_1,j_2)\in \mathcal B}(\Phi_{(j_1,j_2)}+\Phi_{(j_1,j_2)}^{\dagger}).
\end{align} 
At strong coupling i.e. $\lambda\to \infty$, this leads to a theory with a single non-trivial representation corresponding to a chiral majorana fermion with $c_{-}=9/2$. More precisely, the sectors $(0,0),(1,2),(2,1),(3,3)$ form the new vacuum of the theory while the sectors $(1,1),(2,2),(0,3),(3,0)$ are identified into a single fermionic sector. The remaining sectors get confined. The $c_{-}=9/2$ mode can be mapped to single chiral Majorana mode with $c_{-}=1/2$ by symmetric surface pasting of $p+ip$ superconductors described in Sec.~\ref{sec:class}. Therefore by inducing topological order on the surface, it is possible to assemble a pattern of chiral currents that corresponds to the hinge modes obtained from a non-trivial  $\ms{R}_{4z}$ symmetric second-order superconductor. As a corollary one can completely gap out the surface of second-order $\ms{R}_{4z}$ symmetric superconductor by inducing surface topological order.

\section{Boundary-Obstructed Topology with twofold rotation symmetry \texorpdfstring{$\ms C_{2z}$}{}}
\label{sec:botsc}

In this section we study the case where the spatial rotoinversion symmetry is broken down to the $\ms{C}_{2z}$ subgroup. We find that a BdG model with four (modulo eight) Weyl-points and $\ms{C}_{2z}$ symmetry still furnishes a topological superconductor which supports a chiral Majorana hinge mode on its surface. However the mode is no longer protected by the bulk topology and instead is \emph{boundary-obstructed}, in the sense that it can be gapped out by a purely surface deformation.

\subsection{Boundary-obstruction and Wannier representation}
Before discussing the topology of our system with symmetry broken down to $\ms C_{2z}$, we briefly discuss how this symmetry reduction affects the Cooper instability of the system. We still expect the normal state to have the Weyl points on the $k_z = 0, \pi$ planes since they were pinned on the planes by  $\ms C_{2z} \ms T$ symmetry. Additionally, we still expect a minimum of $4$ Weyl points, a pair at $\pm\bm K$ and another at $\pm \bm K^\prime$. Even though the two pairs are not related by any symmetry of the system, we cannot have only a single pair due to the fact that each Weyl-point in a pair related by time-reversal symmetry have the same chirality. This, in conjunction with the Nielsen Ninomiya theorem requires a minimum of two pairs.

\begin{figure}[t]
    \centering
    
    \includegraphics[scale=1]{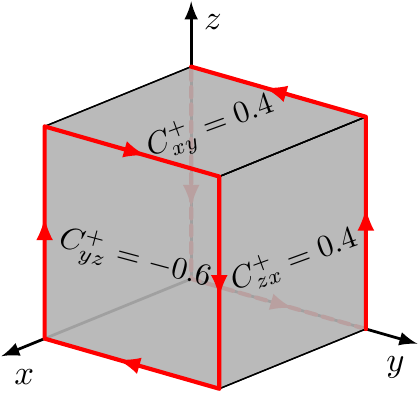}
    \caption{The Majorana zero modes of the model Hamiltonian in Eq.~(\ref{eq:C2z_model}). The surface Chern numbers $C^{\pm}_{ij}$ are as defined in Eq.~(\ref{eq:surface_chern}).}
    \label{fig:majorana_chiral_modes}
\end{figure}


In the absence of the $\ms R_{4z}$ symmetry, one no longer requires $|\Delta_{\bm K}| = |\Delta_{\bm K^\prime}|$. This however does not change the fact that $\Delta_I = d^y_I \mathbb{1}$ still is an eigenmode of the self-consistent equation. Moreover, we still expect a regime in which it is the leading instability as it remains to be the only mode that completely gaps out the Fermi-surfaces of the Weyl semimetal. 

We illustrate boundary-obstructed topology in the $\ms{C}_{2z}$-symmetric case via a specific simplified model, 
\begin{align}\label{eq:C2z_model}
    {H}(\bm k) =& \[\gamma_x + \cos(k_x)\] \sigma_x \tau_z  +\sin(k_z) \sigma_y \tau_z  \nonumber \\ 
    &+ \[\cos(k_y) + \cos(k_z) -1\] \sigma_z \tau_z - \mu \tau_z \nonumber \\
    &+ \sin(k_y) \tau_x + \sin(k_x)  \tau_y.
\end{align}
Numerically solving for the chiral Majorana hinge modes, we obtain the profile shown in Fig.~\ref{fig:majorana_chiral_modes}. The sample has two separate chiral modes that are related by $\ms C_{2z}$ symmetry. These Majorana chiral modes can be removed by for example gluing two $2$D $p+ip$ superconductors with opposite Chern numbers on the two opposite $xz$-surfaces without breaking the symmetry. The model can therefore at best be boundary-obstructed.

From the point of view of bulk Wannier representability, the case with only $\ms C_{2z}$ symmetry is simpler than the case with the more restrictive $\ms R_{4z}$ symmetry. The only restriction of $\ms C_{2z}$ is for the Wannier centers to come in pairs that are related by the symmetry, but otherwise the exact positions can be arbitrary. 

This might seems counter-intuitive at first, since the existence of the chiral modes on the hinges indicate the existence of some sort of a Wannier obstruction. If the bulk is Wannier representable, the only remaining possibility is that the stand-alone surface not be Wannier representable. We discuss this in some detail. The terms in model in Eq.~(\ref{eq:C2z_model}) can be re-organized as
\begin{align}\label{eq:dimerized_limit_ham}
{H}(\bm k) =&\;  H_{p+ip}(\bm k) + H_{\text{SSH}}(\bm k)
\end{align}
with,
\begin{align}
     H_{p+ip}(\bm k)=&\; \[\cos(k_y) + \cos(k_z) -1\] \sigma_z \tau_z  \nonumber \\ &\;+\sin(k_z) \ \sigma_y \tau_z + \sin(k_y) \tau_x,  \nonumber \\
 H_{\text{SSH}}(\bm k) =&\;  \[\gamma_x + \cos(k_x)\] \sigma_x \tau_z + \sin(k_x)  \tau_y,
\end{align}
The $\mc H_{p+ip}(\bm k)$ term describes two $2$-dimensional layers parallel to the $yz$-plane with opposite Chern numbers trivially stacked, while the $\mc H_{\text{SSH}}(\bm k)$ term describes an SSH-like coupling between the layers as shown in Fig.~\ref{fig:chern_stack}. An insulating (i.e without particle-hole symmetry) version of this model is also discussed in \oncite{Khalaf_Benalcazar_Hughes_Queiroz_2019}. Looking at the case when $\gamma_x = 0$, as in Fig.~\ref{fig:chern_stack}, it is clear that the surfaces of the sample (when cut in the $yz$-plane) are not Wannier representable because of the dangling $p+ip$ superconducting layer at each end. Away from the $\gamma_x = 0$ limit the situation is less obvious. However, the Wannier states would evolve smoothly as we move away from the fully dimerized limit, thus the situation would remain unchanged.

\begin{figure}[t]
    \centering
    \includegraphics[scale=1.2]{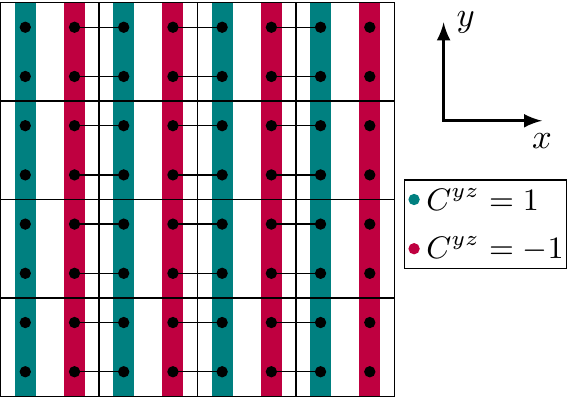}
    \caption{The model in Eq.~(\ref{eq:C2z_model}) can be viewed as the stacking of Chern superconducting layers with SSH like coupling between the layers. In the fully dimerized limit, it is clear that the bulk of the system is Wannier representable, whereas the surfaces perpendicular to the $x$-direction are not.}
    \label{fig:chern_stack}
\end{figure}

\subsection{Defect approach}
We show that the low energy properties of the general Hamiltonian in Eq.~(\ref{eq:general_bgd_ham}) even in the absence of the $\ms R_{4z}$ symmetry leads to a surface theory that is gapped in a topologically non-trivial way, leading to hinge chiral modes. We consider the system with cylindrical hinges along the $z$-directions. We take the radius of the cylinder to be much larger than the inter-atomic distance. The surface theory at each point on the surface of the cylinder can then be taken as that of a straight edge tangent to that point. The rounded hinge can be parametrized by an angle $\theta$ and we define $\hat{\bm n}_{\perp}(\theta)$ as the unit vector perpendicular to the tangent surface, and $\hat{\bm n}_{||}(\theta)$ as the direction parallel to the surface and the $xy$-plane.  Thus at each point on the surface, $\hat{\bm n}_{\perp}(\theta)$, $\hat{\bm n}_{||}(\theta)$, and $\hat{\bm n}_{z})$ constitute and orthonormal coordinate basis. See Fig.~\ref{fig:surface_defect} for an ilustration of the geometry.
\begin{figure}
    \centering
    \includegraphics[scale = 0.9]{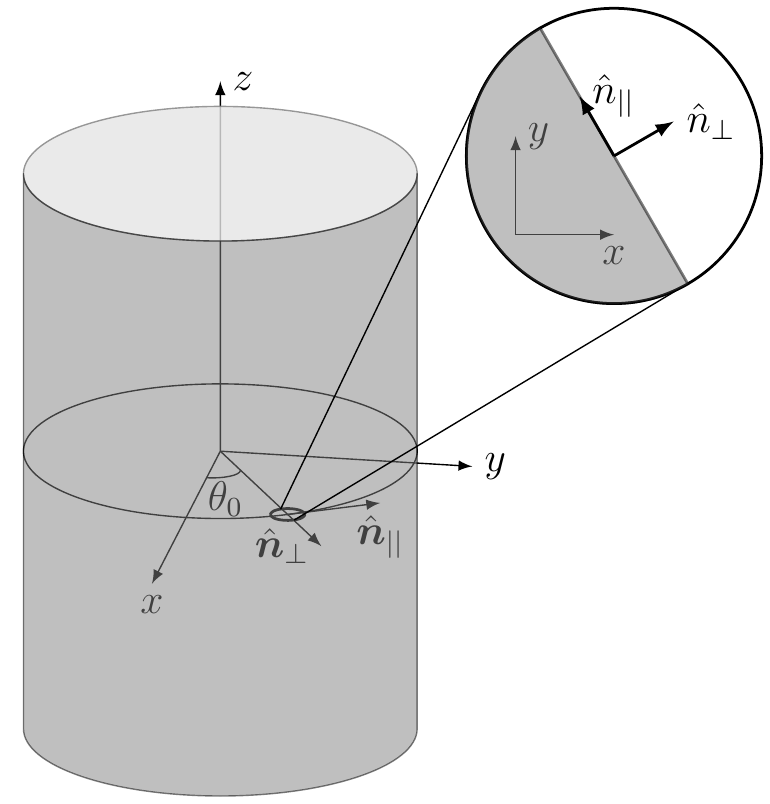}
    \caption{Real space geometry of the sample.}
    \label{fig:surface_defect}
\end{figure}

Since we are interested in the low energetics of the system, we study the system near the Weyl points, and take the order parameter to be small of order $\epsilon$ and write, 
\begin{align}
    \Delta_{1,2}(\bm k) = \epsilon g_{1,2}(\bm k).
\end{align} 

If we start with a particle near the $\bm K$ point, a surface in the $\theta$ direction would \emph{scatter} the particle back, flipping its momentum in the $\hat{\bm n}_{\perp}(\theta)$ direction. Generically, the momentum of this scattered particle will not coincide with another Weyl point. A special case is when $\hat{\bm n}_{\perp}(\theta)$ is in the same direction as $\bm K$, in which the surface mix the momenta at the $\bm K$ point with the $-\bm K$ point. We label such special direction with $\theta_0$. We will reserve the subscripts $||$, $\perp$, and $z$ to indicate the components in the $\hat{\bm n}_{||}(\theta_0)$, $\hat{\bm n}_{\perp}(\theta_0)$, $\hat{\bm n}_{z}$ respectively. 

We expand the Hamiltonian near the Weyl points for a small momentum deviation $\bm q$, and introduce a valley degree of freedom, $\nu_z$, such that $\nu_z = 1$ (respectively $-1$) indicate the $\bm K$ (respectively $-\bm{K}$) point. We define,  $g_i \equiv \eval{g_i(\bm k)}_{\bm k = \bm K}$ and
 \begin{align}
    \vec{\phi}_i \equiv&\; \eval{\frac{\partial f_i(\bm k)}{\partial \vec{k}}}_{\bm k = \bm K}, \quad \vec{\gamma}_i \equiv  \eval{\frac{\partial g_i(\bm k)}{\partial \vec{ k}}}_{\bm k = \bm K}
\end{align}
and set $q_{||} = q_z = 0$, keeping only the first order terms in $\epsilon$ and $q_{\perp}$, and let $q_{\perp} \rightarrow - i \partial_{\perp}$. The resulting Hamiltonian can be written as,
\begin{align}
    H_0 =&\; -i (\phi_{1\perp} \sigma_x + \phi_{3 \perp} \sigma_z) \tau_z \nu_z \partial_{\perp} \nonumber \\ 
    &\; + \epsilon(g_{1} \tau_x + g_{2} \tau_y) \nu_z.
\end{align}
Note that $\phi_{2\perp}=0$ since from Eq.~(\ref{eq:C2zT_constrains}) $f_{2}(\bm k)$ is zero over the entire $k_z = 0$ plane where the Weyl points are located. 

{We  solve  this equation on the half-infinite plane with the vacuum on the $r_\perp>0$  side.   This  equation  has  the  following  zero modes solutions,
\begin{align}
    \psi^\alpha (r_{\perp}) = \chi^\alpha   e^{\Delta_0 r_{\perp}/v_{\perp}},
\end{align}
where we define, 
\begin{align}
    v_{\perp} = \sqrt{\phi^2_{1\perp} + \phi^2_{3\perp} }, && \Delta_0 = \epsilon \sqrt{g^2_{1} + g^2_{2}},
\end{align}
and $\chi^\alpha$ is a eight-component spinor (coming from two band, two valleys, and two Nambu sectors) determined by the following condtions. First, for the zero mode solution to hold, we have 
\be
\tilde{\sigma}_x \tilde{\tau}_y \chi^\alpha = +\chi^\alpha
\label{eq:58}
\ee with
\begin{gather}\label{eq:new_tilde_basis}
    \tilde{\sigma}_x  \equiv \frac{1}{v_{\perp}} (\phi_{1\perp} \sigma_x + \phi_{3 \perp} \sigma_z ), \nonumber  \\  \tilde{\sigma}_y \equiv \sigma_y,
    \qquad \tilde{\sigma}_z  \equiv  i\sigma_y  \tilde{\sigma}_x, \nonumber \\ 
    \tilde{\tau}_x  \equiv \frac{\epsilon}{\Delta_0}  (g_{1} \tau_x + g_{2} \tau_y), \nonumber \\ 
    \tilde{\tau}_z \equiv \tau_z, \qquad  \tilde{\tau}_y \equiv i\tau_z \tilde{\tau}_x.
\end{gather}
Second, the boundary mode is a superposition between incoming and outgoing waves with $\pm \bm K$, or $\nu_z=\pm 1$, depending on the detailed form of the boundary potential. Without loss of generality, in the valley basis, we choose the  condition set by the boundary potential to be 
\be
\nu_x \chi^\alpha = -\chi^\alpha.
\label{eq:59}
\ee
This is equivalent to the boundary condition used in Ref.~\cite{Stone_Roy_2004}. There exist two such eight-component spinors satisfying the above boundary conditions.} 

\begin{figure}[t]
    \centering 
    \subfloat[]{\includegraphics[scale=0.9, trim = 0 0 145 0, clip]{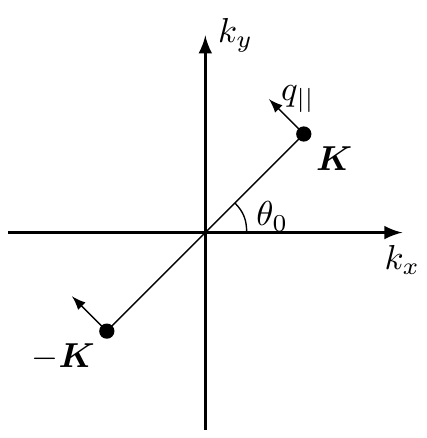}}
    \subfloat[]{\includegraphics[scale=0.9, trim = 140 0 0 0, clip]{small_q_diviations.pdf}}
    \caption{The relative change in momentum between the two valleys for (a) adding momentum to the particles to find the dispersion, (b) changing the direction of the surface by a $\delta \theta$. }
    \label{fig:valley_relative_momentum}
\end{figure}

{Next we find the form of the boundary Hamiltonian for a generic $q_{||}$ and $q_{z}$ and for a generic angular position $\theta=\theta_0+\delta\theta$ on the surface.  For a fixed angular position, the deviation in momenta at the $\bm K$, and $-\bm K$ points has the same direction, see Fig.~\ref{fig:valley_relative_momentum}(a). Upon projecting to the two-dimensional subspace for the boundary states, we get }
\begin{align}
    h(q_{\parallel}, q_{z}) = \hat{P} \[\phi_{2z} \tilde{\sigma}_y  \tilde{\tau}_z q_z + (\beta_1 q_{z} + \beta_2 q_{||})\tilde{\tau}_y\]
\end{align}
where $\hat P$ is the projection onto the subspace and we have defined, 
\begin{align}
    \beta_1 = \frac{\epsilon}{2}  \Tr \tilde{\tau}_y (\gamma_{1 z} \tau_y + \gamma_{2 z} \tau_x), \nonumber \\ 
    \beta_2 = \frac{\epsilon}{2} \Tr \tilde{\tau}_y (\gamma_{1 ||} \tau_y + \gamma_{2 ||} \tau_x).
\end{align}


{For a different surface parameterized by the angle $\theta = \theta_0 + \delta \theta$, the axis of $q_\|=0$ is rotated by $\delta\theta$. In the new coordinate system,  effectively the perturbation incurred are opposite shifts $p_{||} = \pm |K|\delta\theta$ in the positions of Weyl points, shown in Fig.~\ref{fig:valley_relative_momentum}(b). It turns out that the perturbation terms that survives projection onto the two-dimensional subspace is 
\begin{align}
    h(\delta \theta) =\hat P m  \tilde{\sigma}_z \tilde{\tau}_z \delta \theta.
\end{align}
where
\begin{align}
    m = \frac{|K|}{2} \Tr\tilde{\sigma}_z(\phi_{1||} \sigma_x + \phi_{3 ||} \sigma_z).
\end{align}}

Putting the two perturbations together we get a two-band Hamiltonian, 
\begin{align}
    h(q_{\parallel}, q_{z}, \delta \theta) =& \ \hat P  \[\phi_{3z} q_z  \tilde{\sigma}_y \tilde{\tau}_z + (\beta_1 q_{z} + \beta_2 q_{||}) \tilde{\tau}_y \right. \nonumber \\
    &+ \left. m \delta \theta  \tilde{\sigma}_z \tilde{\tau}_z \]
\end{align}
which describes a  2D Dirac fermion with a mass domain wall at $\delta\theta = 0$. Such a Hamiltonian is known to host chiral propagating modes that are localized at the domain wall~\cite{Semenoff_2008, Fosco_1999}. This concludes our proof. 
\subsection{Two-band vs four-band Weyl semimetals}
So far we have restricted out discussion on \emph{two-band} Weyl semimetals -- that is, the four Weyl points are formed by two bands across the full Brillouin zone, which are non-degenerate except at Weyl points. Since there are no Kramers degeneracy at high-symmetry points, necessarily the time-reversal symmetry satisfies $\ms{T}^2=1$.

{In Sec.~\ref{sec:2} we have remarked that the spin-full version of time-reversal symmetry with $\ms{T}^2=-1$ is incompatible with $\ms R_{4z}$ symmetry. However, it is possible to retain only a two-fold rotational symmetry $\ms{C}_{2z} = \ms{R}_{4z}^2$ and have $\ms{T}^2=-1$. Due to the additional Kramer's degeneracy, such a Weyl semimetal involves four bands, given by the following Hamiltonian $H=\int d\bm{k} \psi^\dagger_{\bm k} \mathcal{H}_n \psi_{k}$ where}
\begin{align}
    {H}_n (\bm k) = f_1(\bm{k}) \sigma_x + f_2(\bm k)\sigma_y + f_3(\bm k) \sigma_z + f_3'(\bm k) \sigma_z s_x  -\mu,
    \label{eq:67}
\end{align}
where $s_{z}$ is the Pauli matrix representing an additional spin degree of freedom,  $f_{1,3}(\bm k)$ are even functions and $f'_{3}(\bm k)$ and $f_2(\bm{k})$ are odd. Such a Hamiltonian preserves a time-reveral symmetry $\ms{T}'=i s_y K$ that squares to $-1$. The two-fold rotation symmetry is represented as $\ms{C}_{2z} = is_z$. The location of the Weyl points are given by the conditions
\begin{align}
    f_1(\bm{k})=0,~f_2(\bm k)=0, ~f_3(\bm k)=\pm f_3'(\bm k).
\end{align}
As a concrete example, such a Weyl semimetal with four Weyl points ais realized by the lattice model in which
\begin{align}
    f_1(\bm k)=&\cos k_x + \cos k_y +\cos k_z -2 ,~~f_3(\bm k) = 1/2 \nonumber\\
    f_3'(\bm k) =& \sin k_x ,~~f_2(\bm k) =\sin k_z.
    \label{eq:68}
\end{align}

It is straightforward to show that a $p+ip$ pairing order, e.g., with 
\be
\int d{\bm {k}} \psi^\dagger_{\bm k} [\Delta_x\sin(k_x)+i\Delta_y\sin(k_y)]\sigma_z s_z (\psi^\dagger_{\bm k})^T+h.c.
\label{eq:69}
\ee
gaps out all Fermi surfaces enclosing the Weyl points. 
However, one can readily verify that such a system does \emph{not} host chiral hinge modes, even though the low-energy spectrum \emph{in the bulk} is identical to that of the two-band model. It turns out that the low-energy surface states, which we relied on in the previous subsection to derive the hinge states, in general are \emph{not} solely determined by the low-energy bulk states.  In particular,  having a four-band normal state, the boundary conditions given by Eq.~\eqref{eq:58} and ~\eqref{eq:59} does not reduce the boundary modes to a two-dimensional subspace.

This obstacle can be removed by lifting the $\ms{T}'$ symmetry. This removes all the Kramers degeneracies at high symmetry points and one can separate the four-band model into one with two Weyl bands and two remote bands.
For example, one can include a perturbation from a $\ms{T}'$ breaking, $\ms C_{2z}$ preserving term $\sim M s_z\sigma_z$. As long as $M$ is sufficiently small, it does not affect the band structure near the Weyl points, but it lifts the degeneracy along $k_y=0$. With this term there remains a spinless version of time-reversal symmetry $T=\mathcal K$. Using the argument in the previous subsection, we obtain that in the weak-pairing limit, such a model hosts gapless hinge modes. We indeed confirmed this by numerically solving the lattice model at a finite system size. Unfortunately, however, in general the correct form of the $\ms{T}'$-breaking perturbation that fully disentangles the Weyl bands from remote bands depends on the detailed model and requires a case-by-case analysis.

\section{Conclusion}

{In this work, we have shown that in a time-reversal symmetric doped Weyl semimetal, the combination of symmetry constraints ($\ms{R}_{4z}$ and $\ms T$) and momentum space structure of a finite-range attractive interaction naturally leads to a chiral superconducting state. By analyzing} the topological properties of the superconducting state, we show identify it is a second-order topological phase with chiral Majorana hinge modes {traversing} the surface. 

 We have also analyzed the classification of general BdG Hamiltonians with rotoinversion symmetry supporting second-order topology and found that the classification to be $\mathbb{Z}_2$. We show that the hinge modes can be removed by inducing strong surface interaction leading to a topologically ordered surface state. Crucially such a topologically ordered system with rotoinversion symmetry cannot be realized in strictly two dimensions (i.e without a three dimensional bulk) and is therefore anomalous. The less constrained system with only twofold symmetry is shown to be  boundary-obstructed while also hosting chiral Majorana hinge modes.
 
{In a broader context, Our work showed that the nontrivial topology and gapless excitations in a topological semimetal provide a natural platform for novel topological superconductivity. It will be interesting to explore possible topological superconducting phases from other types of topological semimetals. }

\acknowledgements
We thank Ming-Hao Li, Titus Neupert and Sid Parameswaran for useful discussions. AT acknowledges funding by the European Union’s Horizon 2020 research and innovation program under the Marie Sklodowska Curie grant agreement No 701647. AJ and YW are supported by startup funds at the university of Florida.

\appendix

\section{Calculating the leading Cooper instability.}
\label{app:cooper_leading_instability}
Here we discuss the solutions of the self-consistent equation
\begin{align}
    \Delta_I= T_c \sum_{\omega_m, I^\prime} \int d\delta \bm k
    V_{I I^\prime}P_{I^\prime}(\widehat{\delta \bm k})\frac{ \Tr[P_{I^\prime}(\widehat{\delta \bm k}) \Delta_{I^\prime}]}{\omega^2_m+\xi^2_{I^\prime}(\delta \bm k)}
\end{align}
in more detail.
As discussed in the main text we have, 
\begin{align}
    P_I(\widehat{\delta \bm k}) = \frac{1}{2}\(\mathbb{1} + \hat{\bm n}_I(\widehat{\delta \bm k})\cdot \bm \sigma\)
\end{align}
and,
\begin{align}
    &\Tr (P_I(\widehat{\delta \bm k}) \Delta_I ) = \nonumber  \\ 
    & \qquad i \hat{n}^y_{I}(\widehat{\delta \bm k}) \Delta_I^0 + i d^y_I - d^x_I \hat{n}^z_{I}(\widehat{\delta \bm k}) + d^z_I \hat{n}^x_{I} (\widehat{\delta \bm k}).
\end{align}
From the form of $\hat{\bm n}_I(\widehat{\delta \bm k})$ we see that, 
\begin{align}
    \hat{\bm n}_I(-\widehat{\delta \bm k}) = -\hat{\bm n}_I(\widehat{\delta \bm k}).
\end{align}
Further, using the $\ms C_{2z}\ms T$ symmetry we have that, \begin{align}
    \hat{n}^{x,z}_I(\widehat{\delta k}_x, \widehat{\delta k}_y, -\widehat{\delta k}_z) = \hat{n}^{x,z}_I(\widehat{\delta \bm k}) \\
    \hat{n}^{y}_I(\widehat{\delta k}_x, \widehat{\delta k}_y, -\widehat{\delta k}_z) =- \hat{n}^{y}_I(\widehat{\delta \bm k}).
\end{align}
We conclude from the above equations that terms that are odd in either $\hat{\bm n}_I(\widehat{\delta \bm k})$ or $\hat{n}^{y}_I(\widehat{\delta \bm k})$ will vanish upon integrating over the solid angle.
\begin{align}
    &\Delta_I= i\Delta^0_I \sigma_y +id^y_I \mathbb{1} - d_I^x \sigma_z + d_I^z \sigma_x = \nonumber \\ 
    &\frac{T_c}{2} \sum_{\omega_m, I^\prime} \int d\delta \bm k \frac{V_{II^\prime}}{\omega^2+\xi^2_{I^\prime}(\delta \bm k)} \[ i (\hat{n}^y_{I^\prime}(\widehat{\delta \bm k}))^2 \Delta_{I^\prime}^0 \ \sigma_y + i d^y_{I^\prime} \mathbb{1} \right. \nonumber \\ 
    & \qquad \quad+ \( d^z_{I^\prime} \hat{n}^x_{I^\prime} (\widehat{\delta \bm k}) - d^x_{I^\prime} \hat{n}^z_{I^\prime}(\widehat{\delta \bm k}) \)\hat{n}^x_{I^\prime}(\widehat{\delta \bm k})\sigma_x\nonumber \\
    & \qquad \quad+ \left. \( d^z_{I^\prime} \hat{n}^x_{I^\prime} (\widehat{\delta \bm k}) - d^x_{I^\prime} \hat{n}^z_{I^\prime}(\widehat{\delta \bm k}) \)\hat{n}^z_{I^\prime}(\widehat{\delta \bm k}) \ \sigma_z\].
\end{align}
We see that the both the singlet, and the $\sigma_y$ channel of the triplet pairing form independent solutions of the self-consistent equation. However the $\sigma_x$ and $\sigma_z$ channels do not, they can in general mix together. 
\begin{align}
    &\Delta^0 = \frac{T_c}{2} \sum_{\omega_m, I^\prime} \int d\delta \bm k \frac{V_{I I^\prime}}{\omega^2+\xi^2_{I^\prime}(\delta \bm k)} (\hat{n}^y_{I^\prime}(\widehat{\delta \bm k}))^2\Delta_{I^\prime}^0 \\
    &d^y_I = \frac{T_c}{2} \sum_{\omega_m, I^\prime} \int d\delta \bm k \frac{V_{I I^\prime}}{\omega^2+\xi^2_{I^\prime}(\delta \bm k)} d^y_{I^\prime} \\
    &\mqty(d^x_I \\ d^z_I) = \frac{T_c}{2} \sum_{\omega_m, I^\prime} \int d\delta \bm k \frac{V_{I I^\prime}}{\omega^2+\xi^2_{I^\prime}(\delta \bm k)} \nonumber \\ &\quad\mqty((\hat{n}^z_{I^\prime}(\widehat{\delta \bm k}))^2 && -\hat{n}^x_{I^\prime}(\widehat{\delta \bm k})\hat{n}^z_{I^\prime}(\widehat{\delta \bm k}) \\ 
    -\hat{n}^x_{I^\prime}(\widehat{\delta \bm k})\hat{n}^z_{I^\prime}(\widehat{\delta \bm k}) && (\hat{n}^x_{I^\prime}(\widehat{\delta \bm k}))^2) \mqty(d^x_{I^\prime} \\ d^z_{I^\prime})
\end{align}

Performing the Matsubara sum we have,
\begin{align}
    \sum_{\omega_m} \frac{T_c}{\omega_m^2 + \xi^2_{I^\prime}(\delta \bm k)} = \frac{1}{2\xi_{I^\prime}(\delta \bm k)} \tanh{\frac{\xi_{I^\prime}(\delta \bm k)}{2T_c}}.
\end{align}
\begin{widetext}
Upon doing the change of variables, $d \delta \bm k \rightarrow d\Omega \  d\xi N_{I^\prime}(\xi, \widehat{\delta \bm k}) $, and using
\begin{align}
    \int^\Lambda_{-\Lambda} d\xi \ {N(\xi, \widehat{\delta \bm k})} \frac{1}{2\xi} \tanh{\frac{\xi}{2T_c}} = N(0, \widehat{\delta \bm k})\int^{\Lambda/T_c}_{0} dx \ \frac{1}{x} \tanh{\frac{x}{2}} \approx N(0, \widehat{\delta \bm k})\log\(\frac{\Lambda}{T_c}\).
\end{align}
where $\Lambda$ is an upper cutoff either from the band structure or from the interaction, we get 
\begin{gather}
    \Delta^0_I = \frac{1}{2} \sum_{I^\prime} V_{I I^\prime} \Delta_{I^\prime}^0 \int d\Omega N_{I\prime}(0, \widehat{\delta \bm k}) (\hat{n}^y_{I^\prime}(\widehat{\delta \bm k}))^2 \log(\frac{\Lambda}{T^0_c}) \\
    d^y_I = \frac{1}{2} \sum_{I^\prime} V_{I I^\prime} d^y_{I^\prime} \int d\Omega  N_{I^\prime}(0, \widehat{\delta \bm k}) \log(\frac{\Lambda}{T^y_c}) \\ 
    \mqty(d^x_I \\ d^z_I) = \frac{1}{2} \sum_{I^\prime} V_{I I^\prime} \int d\Omega N_{I^\prime}(0, \widehat{\delta \bm k}) \mqty((\hat{n}^z_{I^\prime}(\widehat{\delta \bm k}))^2 && -\hat{n}^x_{I^\prime}(\widehat{\delta \bm k})\hat{n}^z_{I^\prime}(\widehat{\delta \bm k}) \\ 
    -\hat{n}^x_{I^\prime}(\widehat{\delta \bm k})\hat{n}^z_{I^\prime}(\widehat{\delta \bm k}) && (\hat{n}^x_{I^\prime}(\widehat{\delta \bm k}))^2) \mqty(d^x_{I^\prime} \\ d^z_{I^\prime}) \log(\frac{\Lambda}{T^{xz}_c}).
\end{gather}
In order to simplify the notation we make the following definitions, 
\begin{gather}
    N(0) \equiv \int d\Omega N_{I^\prime}(0, \widehat{\delta \bm k}) \\ 
    \expval*{\hat n_I^i, \hat n_I^j} \equiv \frac{\int d\Omega  N_I(0, \widehat{\delta \bm k}) \hat n_I^i(\widehat{\delta \bm k}) \hat n_I^j(\widehat{\delta \bm k})}{N(0)}.
\end{gather}
Note that $\expval*{\hat n_I^y, \hat n_I^y}$ is constant over all Fermi-surfaces, and $\expval*{\hat n_I^i, \hat n_I^j}$ in general is the same for two opposing Fermi-surfaces. Further, because of the rotoinversion symmetry involved we have $|\Delta_{\ms R_{4z} I}| = |\Delta_{I}|$. For the $s$-wave solution even all the phases are equal across all Fermi-surfaces, whereas for the $p$-wave solutions we have $\Delta_{-I} = -\Delta_{I}$.  
\begin{gather}
    1 = \frac{(V_0 + 2V_1+ V_2)N(0)}{2} \expval*{\hat n_I^y, \hat n_I^y}  \log(\frac{\Lambda}{T^0_c}) \\
    1 = \frac{(V_0 - V_2)N(0)}{2} \log(\frac{\Lambda}{T^y_c}) \\ 
    \mqty(d^x_I \\ d^z_I) = \frac{(V_0 - V_2)N(0)}{2}\log(\frac{\Lambda}{T^{xz}_c})     \mqty(\expval{\hat n_I^z, \hat n_I^z} && -\expval{\hat n_I^z, \hat n_I^x} \\ 
    -\expval{\hat n_I^z, \hat n_I^x} && \expval{\hat n_I^x, \hat n_I^x}) \mqty(d^x_{I} \\ d^z_{I}) .
\end{gather}
The critical temperatures can be read off the above equations as, 
\begin{align}
    T^0_c &= \Lambda \exp[-\frac{2/\expval*{\hat n_I^y, \hat n_I^y}}{(V_0 + 2V_1+ V_2)N(0)}] \label{eq:0_T_c} \\
    T^y_c &= \Lambda \exp[-\frac{2}{(V_0 - V_2)N(0)}] \label{eq:y_T_c} \\
    T^{xz1}_{c} &= \Lambda \exp[-\frac{2/\lambda_1}{(V_0 - V_2)N(0)}] \label{eq:xz1_T_c} \\
    T^{xz2}_c &= \Lambda \exp[-\frac{2/\lambda_2}{(V_0 - V_2)N(0)}] \label{eq:xz2_T_c},
\end{align}
\end{widetext}
where $\lambda_1$ and $\lambda_2$ are the eigenvalues of the matrix, 
\begin{gather}\label{eq:xz_channel_matrix}
    \mqty(\expval{\hat n_I^z, \hat n_I^z} && -\expval{\hat n_I^z, \hat n_I^x} \\ 
    -\expval{\hat n_I^z, \hat n_I^x} && \expval{\hat n_I^x, \hat n_I^x})
\end{gather}
such that $\lambda_2 > \lambda_1$. 

The leading instability of the system is the one that produce the highest critical temperature. We start by comparing the different triplet pairing channels together. By choice we have $T_c^{xz2}>T_c^{xz1}$. What is less trivial is comparing $T^y_c$ with $T^{xz2}$. An upper-bound on $\lambda_2$ can be obtained by replacing the off diagonal terms in Eq.~(\ref{eq:xz_channel_matrix}) by their upper-bound. An upper-bound for $\expval{\hat{n}_I^z, \hat{n}_I^x}$ can be found using the Cauchy-Schwarz inequality, 
\begin{align}
    \expval{\hat n_I^x, \hat n_I^z} \leq \sqrt{\expval{\hat n_I^x, \hat n_I^x}\expval{\hat n_I^z, \hat n_I^z}}.
\end{align}
The charactaristic equation of the resulting matrix is, 
\begin{align}
    \lambda\(\lambda - \expval{\hat n_I^x, \hat n_I^x} - \expval{\hat n_I^z, \hat n_I^z}\) = 0.
\end{align}
Then we have when reacing its upper bound,
\begin{align}
    \lambda_2 =  \expval{\hat n_I^x, \hat n_I^x} + \expval{\hat n_I^z, \hat n_I^z}
\end{align}
{On the other hand we have  $\expval{\hat n_I^x, \hat n_I^x} + \expval{\hat n_I^z, \hat n_I^z}= 1- \expval{\hat n_I^z, \hat n_I^z}<1$, since for around a Weyl point $\hat{\bm n}_I(\delta \bm k)$  points in all possible direction. Therefore we conclude that $\lambda_2<1$, and $T^y_c>T^{xz2}_c$.}

In comparing $T^0_c$ and $T^y_c$ we have two different regimes, 
\begin{gather}
    \frac{V_0 - V_2}{V_0 + 2V_1+ V_2} > \expval*{\hat n_I^y, \hat n_I^y}, \qquad T^y_c > T^0_c
    \label{eq:A30}\\ 
    \frac{V_0 - V_2}{V_0 + 2V_1+ V_2} < \expval*{\hat n_I^y, \hat n_I^y} , \qquad T^y_c < T^0_c. 
\end{gather}
We can expect the $T^y_c > T^0_c$ in the case the interaction is sufficiently long rage. {Indeed, if $V_0$ is the dominant component in the $V$'s, \eqref{eq:A30} always holds.}

It is instructive to see how the calculation is carried in the special case of spherical energy contours. In this case we have,
\begin{align}
    [\phi_I \phi_I^T]^{ij} = v^2 \delta_{ij},
\end{align}
and $N(0,\widehat{\delta \bm k})$ to be constant in $\widehat{\delta \bm k}$.
We thus have, 
\begin{align}
    N(0)=4\pi N_I(0,\widehat{\delta \bm k}),
\end{align}
and,
\begin{align}
    \expval*{\hat{n}^i_I,\hat{n}^j_I } = \frac{1}{3} \delta_{ij}
\end{align}
Using this we can write, 
\begin{gather}
    T^0_c = \Lambda \exp[-\frac{6}{(V_0 + 2V_1+ V_2)N(0)}] \label{eq:0_T_c} \\
    T^y_c = \Lambda \exp[-\frac{2}{(V_0 - V_2)N(0)}] \label{eq:y_T_c} \\
    T^{xz1}_{c} = \Lambda \exp[-\frac{6}{(V_0 - V_2)N(0)}] \label{eq:xz1_T_c} \\
    T^{xz2}_c = \Lambda \exp[-\frac{6}{(V_0 - V_2)N(0)}] \label{eq:xz2_T_c},
\end{gather}
In the spherical Fermi-surfaces case the condition for $T^y_c>T^0_c$ reduces to, 
\begin{gather}
    V_0 > V_1 + V_2.
\end{gather}

\section{A comment on the Wannier spectrum of the model with \texorpdfstring{$\ms R_{4z}$}{}}
The Wannier spectrum come form diagonalizing the Wannier Hamiltonian $\hat \nu_i(\bm k)$ defined through the Wilson loops in the $i$-th direction, 
\begin{align}
    e^{i2\pi\hat\nu_i(\bm k)} \equiv \prod_{n=0}^{L_i-1} \mc P(\bm k + {2\pi n\bm e_i/L_i} ),
\end{align}
where $L_i$ is the system size along $i$-th direction, and $\mc P(\bm k) = \sum_{i=1,2} \dyad{u_i(\bm k)}$ is the projection operator on the occupied states. We note that the operator on the RHS of the above equation acts on a 4-dimensional Hilbert space. However, because of the projection operators involved, it has a 2-dimensional null space, and effectively the Wannier Hamiltonian, $\hat\nu_i(\bm k)$, is 2-dimensional. 

When considering only internal symmetries, the Wannier spectrum in the $i$-th direction share the same topological properties with the surface of the system perpendicular to that direction.~\cite{Fidkowski_Jackson_Klich_2011} However, spatial symmetries can impose vastly different constrains on the surface bands and the Wannier bands, thus leading to different topological features.  Indeed for our case, the $R_{4z}$ symmetry act very  differently on the Wilson loop in the $z$-direction and the surface perpendicular to it. The $R_{4z}$ symmetry maps the top surface of the sample to the bottom surface of the sample, and thus does not put any constrains on the surface spectrum. 

\begin{figure}[t]
    \centering
    \subfloat[][$E(k_x, k_y)$]{\includegraphics[scale = 0.55, trim=10 12 0 20, clip]{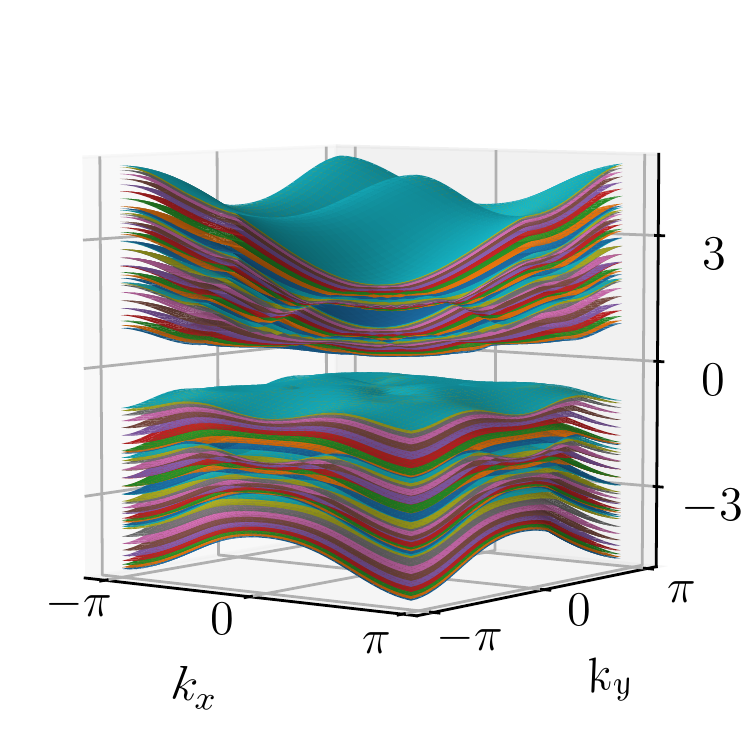}}
    \subfloat[][$\nu_z(k_x, k_y)$]{\includegraphics[scale = 0.55, trim=20 0 0 15, clip]{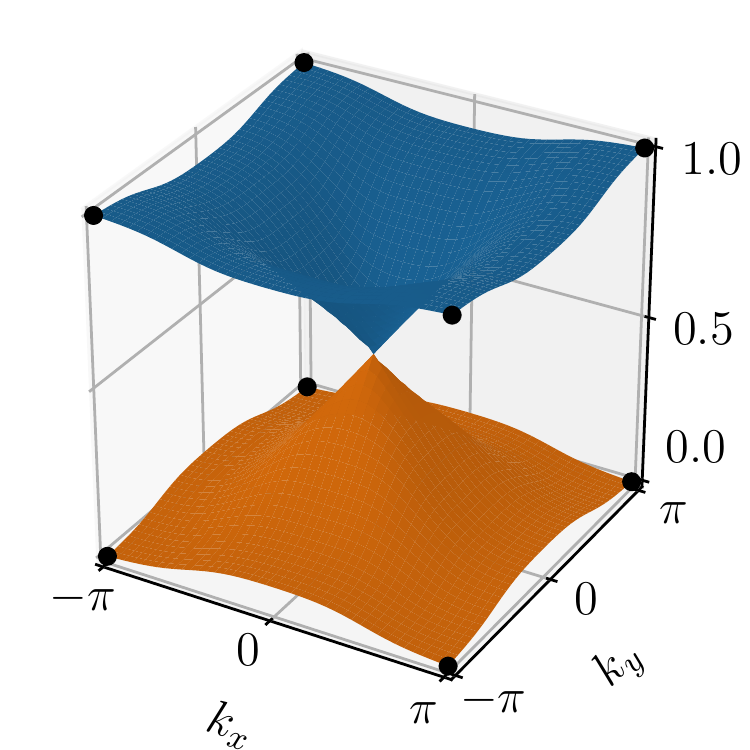}}
    \caption{The energy spectrum (a) with open boundary conditions in the $z$-direction.  Wannier spectrum (b) for the Wilson loops in the $z$-direction. Both graphs are in the topological phase of the system, $\gamma = 0, \Delta = 0.4, \mu = 0.5.$ Size = $(15 \times 15)$. The gaplessness of the Wannier spectrum is protected by $R_{4z}$  while the surface energy spectrum can be gapped without breaking the symmetry.}
    \label{fig:wannier_spec_vs_energy_spec}
\end{figure}

Consider the action of the rotoinversion symmetry on $\mc W_z(\bm k)$ is, 
\begin{align}
    \hat{\ms R}_{4z} \hat{\mc W}_z(\bm k) \hat{\ms R}^{-1}_{4z}  = \hat{\mc W}_z^\dagger(\hat{\ms R}_{4z} \bm k), 
\end{align}
which puts the following constraint on the Wannier spectrum, 
\begin{align}
    \{\nu^i_z (k_x, k_y)\} = \{-\nu^i_z (k_y, -k_x)\}\quad  \text{mod.}\ 1. 
    \label{eq:22}
\end{align}
This action can be thought of as a combination of a chiral symmetry and a fourfold rotation symmetry. In $2$D a chiral symmetry can lead to a symmetry protected Dirac point. We explicitly calculate the Wannier spectrum, and the surface bands for open boundaries in the $z$-direction and compare them. When the chemical potential is zero, we have both spectra to be gapless. However, the gapless mode in the Wannier spectrum is protected by the action of the $R_{4z}$ operator, while gapless mode in the surface spectrum is accidental. Indeed, for non-zero chemical potential, we see that the surface spectrum opens a gap, while the Wannier spectrum does not, see Fig.~\ref{fig:wannier_spec_vs_energy_spec}.

\bibliography{references}
\end{document}